\documentclass[12pt]{article}
\usepackage{amssymb,amsmath,epsfig}
\usepackage{graphicx}
\allowdisplaybreaks
\begin{document}

\title{\bf Anisotropic Tolman V Solutions by Decoupling Approach in $f(R,T^{2})$ Gravity}

\author{M. Sharif \thanks{msharif.math@pu.edu.pk} and Shazmeena
Iltaf \thanks{shazmeenailtaf39@gmail.com}\\
Department of Mathematics, University of the Punjab,\\
Quaid-e-Azam Campus, Lahore-54590, Pakistan.}

\date{}

\maketitle
\begin{abstract}
This paper investigates the behavior of anisotropic static spheres
that are constructed by employing a minimal geometric deformation in
the framework of $f(R,T^{2})$ gravity
($T^{2}=T_{\zeta\nu}T^{\zeta\nu}$, $R$ is the Ricci scalar and
$T_{\zeta\nu}$ is the energy-momentum tensor). We consider a
spherical setup with two sources: seed and additional. It is assumed
that the seed source is isotropic whereas the new source is
responsible for inducing anisotropy. We deform the $g_{rr}$
component to split the field equations into two sets. The first
array corresponds to the isotropic solution whereas the second set
contains the effect of the anisotropic source. The system related to
isotropic source is determined by the metric potentials of Tolman V
solution while three solutions of the second set are constructed
corresponding to three different constraints. The physical
acceptability of all solutions is checked through energy conditions
by employing the radius and mass of PSR J1614-2230 star. We also
examine the stability, mass, compactness and redshift of the
obtained solutions. We conclude that first two solutions satisfy the
viability and stability criteria only for small values of the
decoupling parameter while third solution is stable for its all
possible values.
\end{abstract}
{\bf Keywords:} Gravitational decoupling; Energy-momentum squared gravity; Exact solutions.\\
{\bf PACS:} 04.50.Kd; 04.40.-b; 04.20.Jb

\section{Introduction}

The well-structured universe is composed of celestial bodies whose
physical behavior and arrangement help in comprehending the
evolution of the universe. In this regard, general relativity (GR)
plays an important role in extending our knowledge regarding the
mechanism of cosmos as well as the interior of heavenly bodies. The
solutions of field equations in GR explain different physical
phenomena such as planetary dynamics, black holes and evolution of
the universe. However, the non-linearity of the field equations
creates a problem in the formation of physically viable exact
solutions. Moreover, the extreme central density in compact stellar
objects indicates the presence of anisotropic pressure. Compact
stars have dense cores and their density exceeds the nuclear
density, therefore, pressure must be anisotropic inside them
\cite{15024}. The existence of interacting nuclear matter in dense
compact objects gives rise to the generation of anisotropy
\cite{432}. The inclusion of anisotropy further complicates the
extraction of solutions that effectively describe the structure and
mechanism of celestial bodies. Many researchers have dedicated their
efforts towards devising techniques that aid in the formulation of
well-behaved solutions.

In this regard, Mak and Harko \cite{1} calculated anisotropic
solutions of the field equations and observed that the decreasing
behavior of matter density and pressure, supports the interior of
astronomical objects. Gleiser and Dev \cite{3} analyzed the
stability of the self-gravitating system in the presence of
anisotropic pressure and found that the physical structure of a star
is significantly affected by pressure anisotropy. Sharma and Maharaj
\cite{8000} obtained the exact anisotropic solutions to the field
equations for a linear equation of state. The graphical behavior of
different parameters (mass, energy density and pressure gradient)
have been observed for anisotropic stars by introducing conformal
motion \cite{298}. Kalam et al. \cite{8001} constructed an
anisotropic solution in the framework of Krori and Barua (KB)
metric. The different possibilities have been discussed for the
existence of anisotropic compact star in higher-dimensional
spacetime \cite{8002}. Maurya et al. \cite{8003} computed charged
anisotropic solutions representing compact stars and investigated
various physical properties (energy conditions, stability and
anisotropy).

Recently, the approach of gravitational decoupling through a minimal
geometric deformation (MGD) was introduced which helps in finding
new analytical solutions in astrophysics as well as cosmology.
Ovalle \cite{4} first introduced the decoupling technique in the
context of Randall-Sundrum braneworld. This technique adds a new
gravitational source coupled to the energy-momentum tensor of the
seed matter distribution through a dimensionless parameter. In this
approach, deformation is applied only in the radial metric component
while the temporal metric function remains unaltered. Consequently,
non-linear field equations are split into two sets that exclusively
correspond to seed or additional source. The solution of each set is
evaluated separately and combined afterward to obtain a new
solution. The technique of decoupling has been generally used to
extend the domain of isotropic solutions by incorporating an
anisotropic source.

Ovalle \cite{2*} developed anisotropic solutions through this
technique by using Tolman IV ansatz as a solution of the set
associated with the isotropic source. Hairy black hole solutions
were obtained by incorporating different characteristics of the
additional gravitational source \cite{99}. Anisotropic solutions
analogous to Durgapal-Fuloria \cite{7*}, Buchdahl \cite{01*} and KB
\cite{1508} spacetimes have also been constructed through the MGD
scheme. Morales and Tello-Ortiz \cite{900} studied the charged
anisotropic Heintzmann solution and checked the graphical behavior
of matter variables for different stars. The effect of anisotropy on
the Korkina-Orlyanskii \cite{111} and Tolman VII \cite{88*} perfect
fluid solutions has been observed through decoupling technique.
Casadio et al. \cite{15} applied the gravitational decoupling method
to develop anisotropic solutions by applying the condition of
vanishing complexity factor. Sharif and Sadiq \cite{6} analyzed the
effect of charge on decoupled solutions and concluded that stability
enhances in the presence of charge. Zubair and Azmat \cite{7}
computed an anisotropic solution by specifying the system
corresponding to the isotropic source through Tolman V solution.
Rinc\'{o}n et al. \cite{99*} analyzed the features of
(2+1)-dimensional black hole using the MGD approach. Contreras et
al. \cite{50505} applied decoupling through MGD on the exterior
Schwarzschild solution to formulate new solutions. The gravitational
decoupling method has also been adopted to generate axially
symmetric black hole solutions \cite{101}. Recently,
Carrasco-Hidalgo and Contreras \cite{3000} formulated a new model
for ultra-compact anisotropic star through the MGD method and used
the zero complexity condition to close the system of field
equations.

Cosmologists initially believed that the attractive nature of
gravitational force will ultimately lead to the decelerated
expansion of the cosmos. However, on the contrary, different
observations confirm its current accelerated expansion. Supernova
type Ia has a known brightness, that can be used as standard candles
to determine the luminosity distance of galaxies. The
distance-redshift relation is the most essential approach for
measuring the history of cosmic expansion. One uses observations of
supernova type Ia to calculate luminosity distances at various
redshifts and the other uses measurements of baryon acoustic
oscillations to determine angular diameter distances at several
redshifts. The non-linear feature of luminosity distance versus
red-shift fits with the accelerated expansion of the universe. In
the light of cosmological observations, it is necessary to look for
a new approach that is compatible with our findings. In order to
explain the present accelerated expansion of the universe,
physicists coined the term dark energy to explain the mysterious
force whose negative pressure is causing the galaxies to drift apart
with acceleration. Many efforts have been done by several
researchers to uncover this mysterious force. In this regard, the
cosmological constant is considered as the best candidate to explain
the mysterious aspects of dark energy, but it has two major problems
involving the fine tuning and coincidence problem.

Modified theories have successfully played the role of alternatives
to the dark energy. Such theories are formed by adding the
higher-order curvature invariants in the geometric part of the
Einstein-Hilbert (EH) action. In this regard, the $f(R)$ gravity is
obtained by replacing the Ricci scalar $(R)$ with the generic
function $f(R)$ in the geometric part of EH action \cite{7779}. The
$f(R)$ gravity has been further generalized by introducing coupling
between the matter sector and the geometrical quantities leading to
$f(R,L_{m})$ theory ($L_{m}$ is matter Lagrangian) \cite{38}. The
$f(R,\textbf{T})$ gravity $(\textbf{T}=T_{\zeta\nu}g^{\zeta\nu})$ is
one of such modifications obtained by replacing the matter
lagrangian with the trace of energy-momentum tensor in $f(R,L_{m})$
theory \cite{605}. A generic theory in which matter and geometry are
non-minimally coupled was introduced in \cite{234}, referred to as
$f(R,\textbf{T},R_{\zeta\nu}T^{\zeta\nu})$, where $R_{\zeta\nu}$ is
the Ricci tensor and $T^{\zeta\nu}$ is the energy-momentum tensor.
Sharif and Waseem \cite{3*} constructed the charged anisotropic
solutions and observed that the presence of charge provides more
stable behavior in $f(R)$ theory. Anisotropic solutions by
decoupling technique have been constructed by using isotropic KB
solution in this theory \cite{14*}. Plenty of work has been done to
obtain viable and stable solutions through decoupling technique in
$f(R,\textbf{T})$ and $f(R)$ theories \cite{9903}-\cite{9905}.
Various viable and stable solutions have been obtained through
decoupling technique in $f(R,\textbf{T},R_{\zeta\nu}T^{\zeta\nu})$
\cite{111**,111*} framework.

The Big Bang theory is still the most generally recognized
cosmological model that explains evolution of the cosmos. The theory
specifies how the cosmos evolved from a high-temperature state, a
high density as well as a variety of well-known observations such as
cosmic microwave background radiations, large-scale structures, and
the abundance of light elements. According to the Big Bang theory,
all the matter in the cosmos exploded from a singularity. The
occurrence of singularities is the major issue in GR. The existence
of a singularity at the beginning of the cosmos is explained at high
energy levels where GR is not viable. In this regard, $f(R,T^{2})$
(where $R$ is the Ricci scalar and $T^{2}=T^{\zeta\nu}T_{\zeta\nu}$)
gravity is considered as a favourable theory which resolves the
singularity problem. In 2014, Katirci and Kavuk \cite{1500}
developed this theory by adding a non-linear term
$T^{2}=T^{\zeta\nu}T_{\zeta\nu}$ in the geometric part of the EH
action and discussed the cosmological applications for the model
$f(R,T^{2})=R+\beta T^{2}$, where $\beta$ is the real-valued
coupling parameter. This theory is also referred to as the
energy-momentum squared (EMSG) gravity. Such matter-geometry
coupling explains different eras of the universe as well as the
rotation curves of galaxies. Faraji et al. \cite{16*} used
Friedmann-Lemaitre-Robertson-Walker (FLRW) metric to observe the
effect of coupling parameter and found that the range $(0,2.1 \times
10^{5}]$ provides observationally viable inflationary model. Several
remarkable results have also been obtained by using the Kuchowicz
metric in this scenario \cite{2}. A gravitational model acts only as
an alternative to the mysterious dark energy. These models also
explain the cosmic expansion. In the study of the early universe,
this theory supports a regular bounce, i.e., a cosmos with finite
maximum energy density and a minimal scale factor. It can resolve
the singularity problem with classical prescription.

Using a specific model of EMSG, Board and Barrow \cite{55} computed
exact solutions to discuss different scenarios related to isotropic
distribution, the existence of singularities as well as evolution of
the cosmos. Nari and Roshan \cite{17'} compared the results of GR
and EMSG while describing compact stars. The compactness of a
neutron star has been computed for different values of the coupling
parameter lying within the range $(-10^{-38}cm^{3}/erg$,
$10^{-37}cm^{3}/erg)$ \cite{200}. Bahamonde et al. \cite{71}
discussed the phase space and stability analysis for different
models in EMSG. The viability of isotropic cosmos has also been
analyzed in this theory by using flat FLRW metric \cite{78}. Sharif
and Gul \cite{13*} constructed various cosmological as well as
astrophysical solutions via different techniques and discussed their
viability and stability. Recently, Sharif and Naz \cite{ss} have
computed gravastar structure in the perspective of curvature-matter
coupled theory and concluded that the physical features of gravastar
follow the increasing trend against the thickness of the shell.

In this paper, we decouple the field equations through a deformation
in the radial metric function to generate anisotropic analogs of
Tolman V in the background of EMSG. The paper is organized according
to the following arrangement. In section \textbf{2}, we derive the
field equations for a static sphere filled with two sources which
are decoupled through the MGD scheme in section \textbf{3}. We
develop three anisotropic solutions and use the junction conditions
to specify the unknown constraints in section \textbf{4}. Section
\textbf{5} discusses the physical behavior, viability and stability
of all the solutions graphically. We conclude our results in section
\textbf{6}.

\section{\textbf{$f(R, T^{2})$} Field Equations}

In this section, we construct the field equations in the framework
of EMSG. The complete action for this theory in the presence of additional source is
\begin{equation}\label{1}
S=\frac{1}{2\kappa^{2}}\int f(R,T^{2})\sqrt{-g}d^{4}x+\int
L_m\sqrt{-g}d^{4}x +\alpha\int L_{\Phi}\sqrt{-g}d^{4}x,
\end{equation}
where $\kappa^{2}=\frac{8\pi G}{c^{4}}$, $R$, $L_{\Phi}$ and $L_{m}$
denote the coupling constant, Ricci scalar, Lagrangian densities of
new source $(\Phi_{\zeta\nu})$ and matter distribution,
respectively. Further, $g$ is the determinant of the metric tensor
and $\alpha$ is the decoupling parameter. For the sake of
simplicity, we take the gravitational constant $(G)$ and speed of
light $(c)$ as $1$ and thus $\kappa^2=8\pi$. This action represents
that this theory has additional degrees of freedom. As a result of
the extra force and a matter-dominated era, it is believed that some
important results will be gained to analyze the present cosmic
problems in this gravity. Variation of the action (\ref{1}) with
respect to the metric tensor leads to the following field equations
\begin{eqnarray}\label{2*}
R_{\zeta\nu}f_R+g_{\zeta\nu}\Box
f_R-\nabla_{\zeta}\nabla_{\nu}f_R-\frac{1}{2}g_{\zeta\nu}f
=8\pi\left(T_{\zeta\nu}+\alpha
\Phi_{\zeta\nu}\right)-\Theta_{\zeta\nu}f_{T^{2}},
\end{eqnarray}
where $\Box$ is the d'Alembertian operator (which is defined as
$\Box=\nabla_{\zeta}\nabla^{\zeta}$),
$f_{T^{2}}=\frac{\partial{f}}{\partial{T^{2}}}$,
$f_R=\frac{\partial{f}}{\partial{R}}$ and
\begin{equation}\label{004}
T_{\zeta\nu}=g_{\zeta\nu}L_{m}-2\frac{\partial L_{m}}{\partial
g^{\zeta\nu}}.
\end{equation}
and
\begin{equation}\label{3}
\Theta_{\zeta\nu}=\frac{\partial T^{2}}{\partial
g^{\zeta\nu}}=-2L_m(T_{\zeta\nu}-\frac{1}{2}g_{\zeta\nu}T)
-4\frac{\partial^{2}{L_m}}{\partial{g^{\zeta\nu}}{\partial{g^{\alpha\beta}}}}
T^{\alpha\beta}-TT_{\zeta\nu}+2T^{\beta}_{\zeta}T_{\nu\beta}.
\end{equation}
Different choices of the matter Lagrangian for a perfect fluid have
been analyzed in literature. These include $L_{m}=\pm P$ \cite{23},
$L_{m}=T$ \cite{90} and $L_{m}=\pm\rho$ \cite{67}. The sign depends
on the signatures of the chosen metric. Here we take metric
signatures $(+,-,-,-)$ along with $L_{m}=-P$, which is one of the
probable Lagrangian for a perfect fluid mentioned in the literature
\cite{98,6543}. The energy-momentum tensor for perfect fluid is
\begin{equation}\label{4}
T_{\zeta\nu}=(\rho+P)U_{\zeta}U_{\nu}-g_{\zeta\nu}P,
\end{equation}
where $U_{\zeta}$, $\rho$ and $P$ represent the four-velocity,
energy density and pressure, respectively. Inserting $L_{m}=-P$ in
Eq.(\ref{3}) yields
\begin{eqnarray}\label{5}
\Theta_{\zeta\nu}&=&(3P^{2}+\rho^{2}+4P\rho)U_{\zeta}U_{\nu}.
\end{eqnarray}
By rearranging Eq.(\ref{2*}), we obtain
\begin{equation}\label{8}
G_{\zeta\nu}=\frac{1}{f_R}\Big(8\pi(T^{(m)}_{\zeta\nu}+\alpha\Phi_{\zeta\nu})-g_{\zeta\nu}\Box
f_{R}+\nabla_{\zeta}\nabla_{\nu}f_{R}
-\Theta_{\zeta\nu}f_{T^{2}}+\frac{1}{2}g_{\zeta\nu}(f-R f_{R})\Big),
\end{equation}
where $G_{\zeta\nu}=R_{\zeta\nu}-\frac{1}{2}Rg_{\zeta\nu}$ is the
Einstein tensor.

To describe the spherical structure of self-gravitating objects, we
consider the static line element
\begin{equation}\label{10}
ds^{2}=e^{\xi}dt^{2}-e^{\eta}dr^{2}-r^{2}d\theta^{2}-r^{2}\sin^{2}\theta
d\phi^{2},
\end{equation}
where the metric potentials $(\xi,~\eta)$ depend on the radial
coordinate only.  We consider the model $f(R,T^{2})=R+\beta T^{2}$
to make our results meaningful. Using Eqs.(\ref{4}) and (\ref{5}) in
(\ref{8}), the field equations corresponding to the above model
become
\begin{eqnarray}\label{11}
8\pi(\tilde{\rho}+\alpha\Phi^{0}_{0})&=&e^{-\eta}(\frac{\eta^{'}}{r}-\frac{1}{r^{2}})+\frac{1}{r^{2}},\\\label{12}
8\pi(\tilde{P}-\alpha\Phi^{1}_{1})&=&e^{-\eta}(\frac{\xi^{'}}{r}+\frac{1}{r^{2}})-\frac{1}{r^{2}},\\\label{13}
8\pi(\tilde{P}-\alpha\Phi^{2}_{2})&=&\frac{e^{-\eta}}{4}(2\xi^{''}-\frac{2\eta^{'}}{r}+\frac{2\xi^{'}}{r}-
\xi^{'}\eta^{'}+\xi^{'2}),
\end{eqnarray}
where
\begin{eqnarray}\label{14}
\tilde{\rho}&=&\rho-\frac{1}{16\pi}(\rho^{2}+3P^{2}+8\rho
P)\beta,\\\label{15}
\tilde{P}&=&P-\frac{1}{16\pi}(\rho^{2}+3P^{2})\beta.
\end{eqnarray}
Here, prime represents the derivative with respect to the radial
coordinate and $\beta$ is the model parameter having dimensions of
$L^{2}$. The energy-momentum conservation law does not hold in EMSG
as the covariant derivative of Eq.(\ref{8}) takes the following form
\begin{eqnarray}\nonumber
\nabla^{\zeta}(\Theta_{\zeta\nu}f_{T^{2}})-\frac{1}{2}g_{\zeta\nu}\nabla^{\zeta}f&=&
\tilde P^{'}+\frac{\xi^{'}}{2}(\tilde\rho+\tilde
P)+\alpha\left(\frac{\xi^{'}}{2}(\Phi^{0}_{0}-\Phi^{1}_{1})\right.\\\label{16}&+&
\left.\frac{2}{r}(\Phi^{2}_{2}-\Phi^{1}_{1})-\Phi^{1'}_{1}\right).
\end{eqnarray}
The field equations are non-linear differential equations containing
seven unknown functions: the matter variables $(\tilde{\rho},
~\tilde{P})$, the metric potentials $(\xi,\eta)$ and the extra
source components $(\Phi^{0}_{0},~\Phi^{1}_{1},~\Phi^{2}_{2})$. The
effective energy density, effective radial and tangential pressures
are, respectively, defined as
\begin{eqnarray}\label{17}
\bar{\rho}&=&\tilde{\rho}+\alpha\Phi^{0}_{0},\\\label{18}
\bar{P_{r}}&=&\tilde{P}-\alpha\Phi^{1}_{1},\\\label{19}
\bar{P_{t}}&=&\tilde{P}-\alpha\Phi^{2}_{2}.
\end{eqnarray}
Moreover, the anisotropy generated by $\Phi_{\zeta\nu}$ in the
self-gravitating object has the following form for
$\Phi^{1}_{1}\neq\Phi^{2}_{2}$
\begin{equation}\label{20}
\Delta=\bar{P_{t}}-\bar{P_{r}}=\alpha(\Phi^{1}_{1}-\Phi^{2}_{2}).
\end{equation}
If $\beta\rightarrow0$, the field equations in $f(R, T^{2})$ gravity
will reduce to GR.

\section{Gravitational Decoupling}

In this section, we use the method of gravitational decoupling
through MGD to solve the system of field equations. In order to
incorporate the impact of the decoupling parameter in the perfect
fluid distribution, we consider the following geometric
modifications
\begin{eqnarray}\label{23}
\zeta\rightarrow\xi&=&\zeta+\alpha s(r),\\\label{24} \nu\rightarrow
e^{-\eta}&=&\nu+\alpha h(r),
\end{eqnarray}
where $s$ and $h$ represent the deformations in temporal and radial
components, respectively. As we are using the MGD approach, we
perform geometric deformation only in the radial component while the
temporal function remains unaltered, i.e., $s=0$. We obtain two sets
of equations by using Eq.(\ref{24}) in the set of field equations.
The first set of equations is obtained for $\alpha=0$ and is given
as
\begin{eqnarray}\label{27}
8\pi\tilde{\rho}&=&\frac{1}{r^{2}}-(\frac{\nu}{r^{2}}+\frac{\nu^{'}}{r})=8\pi\rho-
\frac{1}{2}(\rho^{2}+3P^{2}+8\rho P)\beta, \\\label{28}
8\pi\tilde{P}&=&-\frac{1}{r^{2}}+\frac{\nu}{r}(\frac{1}{r}+\xi^{'})=8\pi
P-\frac{1}{2} (\rho^{2}+3P^{2})\beta,\\\label{29}
8\pi\tilde{P}&=&\frac{\nu}{2r}(\xi^{''}r+\frac{\xi^{'2}r}{2}+\xi^{'})+\frac{\nu^{'}}{2}(\frac{\xi^{'}}{2}+\frac{1}{r})=
8\pi P-\frac{1}{2}(\rho^{2}+3P^{2})\beta,
\end{eqnarray}
whereas the second set describing the anisotropic source is given as
\begin{eqnarray}\label{30}
8\pi\Phi^{0}_{0}&=&-\frac{h^{'}}{r}-\frac{h}{r^{2}},\\\label{31}
8\pi\Phi^{1}_{1}&=&-\frac{h}{r}(\frac{1}{r}+\xi^{'}),\\\label{32}
8\pi\Phi^{2}_{2}&=&-\frac{2h}{r}(\xi^{''}+\frac{\xi^{'2}}{2}+\frac{\xi^{'}}{r})-\frac{h^{'}}{2r}
(\frac{\xi^{'}r}{2}+1).
\end{eqnarray}
The energy-momentum conservation equations for both scenarios take
the following forms
\begin{eqnarray}\label{33}
\nabla^{\zeta}(\Theta_{\zeta\nu}f_{T^{2}})-\frac{1}{2}g_{\zeta\nu}\nabla^{\zeta}f&=&\tilde{P^{'}}+
\frac{\eta^{'}}{2}(\tilde{\rho}+\tilde{P}), \\\label{34}
\Phi^{1'}_{1}-\frac{\eta^{'}}{2}(\Phi^{0}_{0}-\Phi^{1}_{1})-\frac{2}{r}(\Phi^{2}_{2}-\Phi^{1}_{1})
&=&0.
\end{eqnarray}
Simultaneously solving Eqs.(\ref{27})-(\ref{29}) provides
complicated expressions of energy density and pressure. To simplify
calculations, we apply the following relation
\begin{equation}\label{35}
P=\rho=\frac{m}{V},
\end{equation}
where $V=\frac{4}{3}\pi r^{3}$ is the volume of sphere. Using this
equation, $\tilde\rho$ and $\tilde P$ can be written in the form of
mass as
\begin{eqnarray}\label{37}
\tilde\rho&=&\frac{3m(16\pi^{2}r^{3}-9m\beta)}{64\pi^{3}r^{6}},\\\label{38}
\tilde P&=&\frac{48m\pi^{2}r^{3}-9m^{2}\beta}{64\pi^{3}r^{6}}.
\end{eqnarray}
The effective energy density and pressure of the isotropic source is
described by Eqs.(\ref{37}) and (\ref{38}), respectively.

\section{Anisotropic Solutions}

Tolman \cite{5*} presented eight perfect fluid solutions of the
field equations in the background of a static spherical spacetime.
We choose the components of Tolman V solution to solve
Eqs.(\ref{27})-(\ref{29}). Tolman V is a natural solution used for
investigating the fluid spheres with infinite density and pressure
at the center. This solution has been previously used to specify the
metric components related to the seed source \cite{7}. The metric
coefficients of Tolman V solution are
\begin{eqnarray}\label{51}
e^{\xi(r)}&=&C^{2}r^{2n},\\\label{52}
e^{\eta(r)}&=&\nu^{-1}=\frac{1+2n-n^{2}}{1-(1+2n-n^{2})(\frac{r}{F})^{W}},
\end{eqnarray}
where $W=\frac{2(1+2n-n^{2})}{1+n}$. The values of the constants
appearing in the metric potentials of Tolman V solution are
determined by using the matching conditions \cite{09}. These
conditions play an important role in the study of stellar geometry
as they provide a smooth junction between the exterior and interior
regions at the boundary $(\Sigma: r=\mathcal{R})$ of the star. The
essential conditions are given as
\begin{equation}\label{101}
(ds^{2}_{+})_{\Sigma}=(ds^{2}_{-})_{\Sigma}, \quad
\bar{P}_{r}(\mathcal{R})=0,
\end{equation}
which imply that radial pressure drops to zero at the boundary which
is essential for the stability of a star. We use
Schwarzschild metric to describe the exterior geometry as
\begin{eqnarray}\label{59*}
ds^{2}_{+}=-\left(\frac{2M}{\mathcal{R}}-1\right)dt^{2}+\left(\frac{2M}{\mathcal{R}}-1\right)^{-1}dr^{2}
-r^{2}d\theta^{2}-r^{2}\sin^{2}\theta d\phi^{2},
\end{eqnarray}
where $M$ represents the mass of compact object. The constants are
evaluated by using the junction conditions as
\begin{eqnarray}\label{59}
C^{2}&=&\frac{M}{n\mathcal{R}^{2n+1}},\\\label{60}
n&=&\frac{M}{\mathcal{R}-2M},\\\label{61}
\left(\frac{\mathcal{R}}{F}\right)^{W}&=&\frac{M(M\mathcal{R}-2M^{2})}{\mathcal{R}(\mathcal{R}^{2}-2M\mathcal{R}-M^{2})}.
\end{eqnarray}
We have three field equations and four unknowns in the second set
comprising additional source. To minimize the number of unknowns
$(h(r),~\Phi^{0}_{0}, ~\Phi^{1}_{1},~\Phi^{2}_{2})$, we need a some
constraint. For this purpose, we use mimic constraints as well as an
equation of state on components of $\Phi_{\zeta\nu}$ to obtain three
viable solutions.

\subsection{Solution I}

To find the values of unknowns, we apply a condition on
$\Phi^{0}_{0}$. For this motive, we use the constraint \cite{14*}
\begin{equation}\label{62}
\Phi^{0}_{0}=\tilde\rho.
\end{equation}
By using Eqs.(\ref{27}), (\ref{30}), (\ref{51}) and (\ref{52}), we
obtain the following expression of the deformation function,
\begin{eqnarray}\nonumber
h&=&-\frac{8\pi\left(n^{2}-2n+(n^{2}-2n-1)\left(\frac{r}{F}\right)^{W}\right)}{
(n^{2}-2n-1)}+\frac{d}{r},
\end{eqnarray}
where $d$ is an integration constant. If $r=0$, then the above
result diverge. So, we assume $d=0$ and obtain the reduced form of
the deformation function as
\begin{eqnarray}\label{656}
h&=&-\frac{8\pi\left(n^{2}-2n+(n^{2}-2n-1)\left(\frac{r}{F}\right)^{W}\right)}{
(n^{2}-2n-1)}.
\end{eqnarray}
Using Eq.(\ref{656}), we calculate the values of the matter
variables as
\begin{eqnarray}\nonumber
\bar{\rho}&=&\frac{\Bigg(\left(\frac{r}{F}\right)^{W}(1+W)(n^{2}-1-2n)
-2n+n^{2}\Bigg)\alpha}{(n^{2}-1-2n)r^{2}}\\\label{908}&+&\frac{3m(16\pi^{2}r^{3}
-9m\beta)}{64\pi^{3}r^{6}},\\\nonumber \bar{P_{r}}&=&-\frac{(1 +
2n)\Bigg(\left(\frac{r}{F}\right)^{W}(n^{2}-2n-1)+n^{2}-2n\Bigg)\alpha}{(n^{2}
-2n-1)r^{2}}\\\label{987}&+&\frac{48m\pi^{2}r^{3}-9m^{2}\beta}{64\pi^{3}r^{6}},
\\\nonumber
\bar{P_{t}}&=&\frac{32r^{4}\left(\frac{r}{F}\right)^{W}W(1+3n+n^{2}-
n^{3})-\left(\frac{r}{F}\right)^{W}(64r^{4}n^{4}-128r^{4}n^{3})}
{64r^{6}(n^{2}-2n-1)}\alpha\\\label{009*}&+&\frac{64r^{4}n^{2\left(\frac{r}{F}\right)^{W}
}-64r^{4}n^{4}+128r^{4}n^{3}}{64r^{6}
(n^{2}-2n-1)}\alpha+\frac{48m\pi^{2}
r^{3}-9m^{2}\beta}{64r^{6}\pi^{3}}.
\end{eqnarray}
We graphically analyze the physical features for PSR J1614-2230 star
with mass $1.97M_{\bigodot}$ ($M_{\bigodot}$ is solar mass) and
radius $11.29km$. The model parameter in this case can be positive
or negative. However, we do not get viable and stable results for
its positive values and thus, we use negative values of $\beta$ to
obtain acceptable solutions. Moreover, the value of the model
parameter is fixed as $-0.1$ for $\alpha=0.1,~0.21$. The energy
density and pressures of a viable model attain maximum values near
$r=0$ and decrease monotonically towards the surface. Moreover,
positive anisotropy, obtained for $P_{t}>P_{r}$, implies the
existence of an outward repulsive force. Figure \textbf{1} indicates
that the physical parameters and anisotropy attain the maximum
values near the center and remain finite throughout for the assumed
values of parameters. Moreover, the radial pressure drops to zero as
$r\rightarrow \mathcal{R}$.
\begin{figure}\center
\epsfig{file=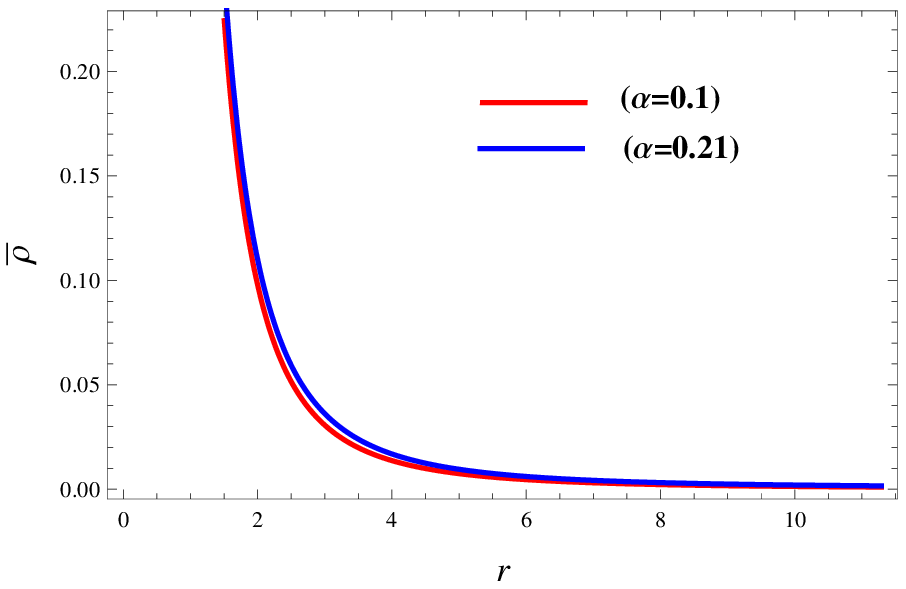,width=0.45\linewidth}\epsfig{file=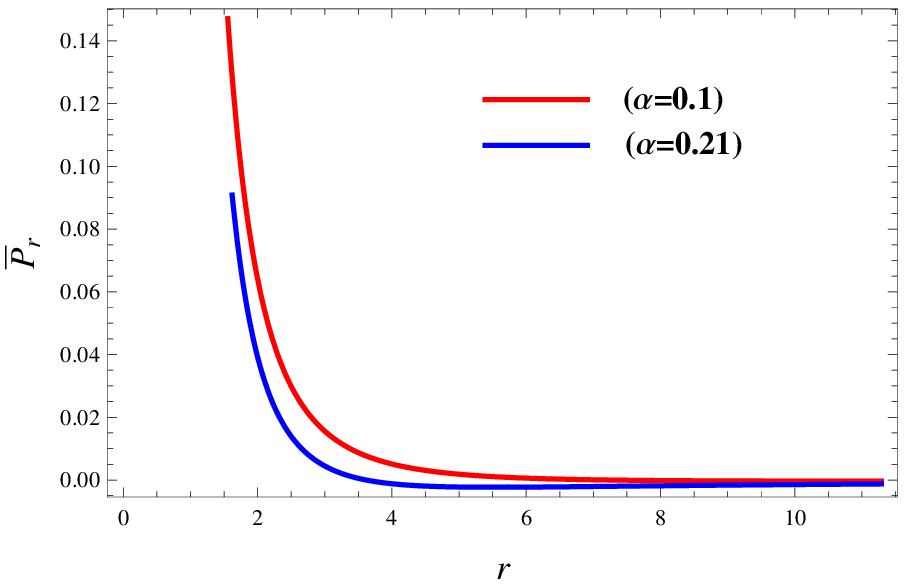,width=0.45\linewidth}
\epsfig{file=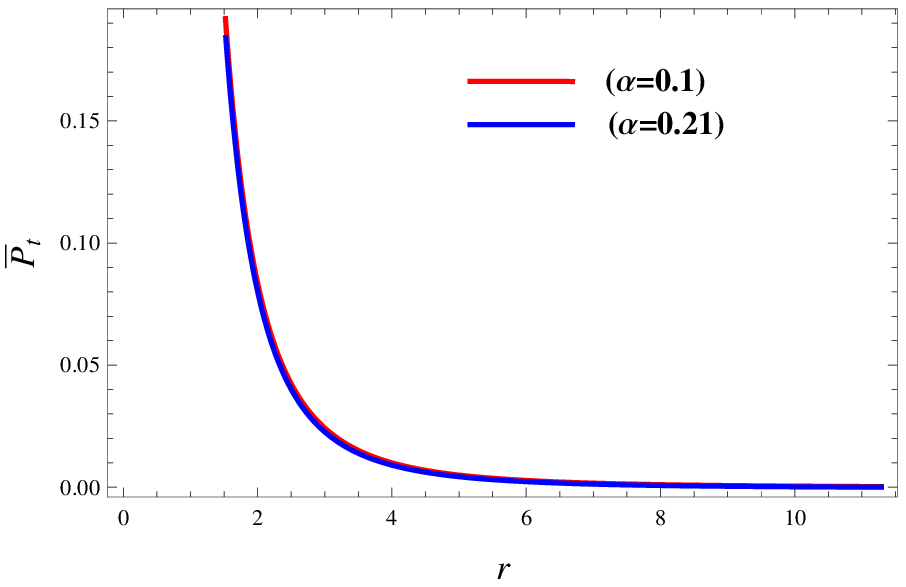,width=0.45\linewidth}\epsfig{file=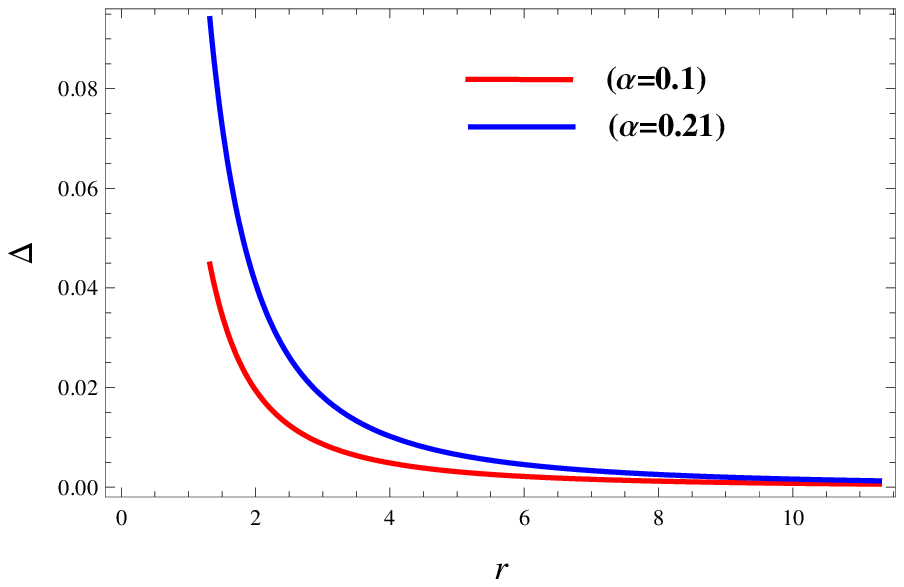,width=0.45\linewidth}
\caption{Behavior of energy density, radial/tangential pressures and
anisotropy of the first solution.}
\end{figure}

\subsection{Solution II}

In this section, we apply a condition on $\Phi^{1}_{1}$ to determine
the deformation function $h$. For this purpose, we use the mimic
constraint \cite{99}
\begin{equation}\label{75}
\Phi^{1}_{1}=\tilde P.
\end{equation}
Using Eqs.(\ref{28}) and (\ref{31}) alongwith the metric potential
of Tolman V solution, the deformation function is evaluated as
\begin{equation}\label{766}
h(r) =
\frac{8\pi\left(n^{2}+(2n^{3}-3n^{2}-4n-1)\left(\frac{r}{F}\right)^{W}\right)}
{(1+2n)(n^{2}-2n-1)}.
\end{equation}
Equations (\ref{52}) and (\ref{766}) provide the minimally deformed
Tolman V solution. Employing Eq.(\ref{766}), we develop the
expressions of effective parameters as
\begin{eqnarray}\nonumber
\bar{\rho}&=&\frac{\left(\frac{r}{F}\right)^{W}(1+W)(1+4n+3n^{2}-2n^{3})-n^{2}}
{(1+2n)(n^{2}-2n-1)r^{2}}\alpha\\\label{5686}&+&\frac{3m(16\pi^{2}r^{3}-9m\beta)}{64\pi^{3}r^{6}},\\\label{080}
\bar{P_{r}}&=&\frac{\left(\frac{r}{F}\right)^{W}(1+4n+3n^{2}-2n^{3})-n^{2}}
{(n^{2}-2n-1)r^{2}}\alpha+\frac{48m\pi^{2}r^{3}-9m^{2}\beta}{64\pi^{3}r^{6}},\\\nonumber
\bar{P_{t}}&=&-\frac{32r^{4}\left(\frac{r}{F}\right)^{W}(2n^{2}+8n^{3}+6n^{4}-4n^{5})
+(5n+1+7n^{2})\left(\frac{r}{F}\right)^{W}W}{64r^{6}(1+2n)(n^{2}-1-2n)}\alpha
\\\nonumber&-&\frac{(2n^{4}-n^{3})W\left(\frac{r}{F}\right)^{W}}{64r^{6}(2n+1)(n^{2}-1-2n)}\alpha
-\frac{2n^{4}}{64r^{6}(1+2n)(n^{2}-2n-1)}\alpha\\\label{4568}&+&\frac{48m\pi^{2}r^{3}
-9m^{2}\beta}{64r^{6}\pi^{3}}.
\end{eqnarray}
We consider the same value of $\beta$ by considering the same star
for $\alpha=0.1,~0.59$. The effective energy density and effective
pressures of a viable model attain maximum values near the center
and decrease monotonically towards the boundary. Figure \textbf{2}
indicates that the effective parameters attain the maximum values
near the core and decrease towards the boundary. Figure \textbf{2}
also describes the positive anisotropy.
\begin{figure}\center
\epsfig{file=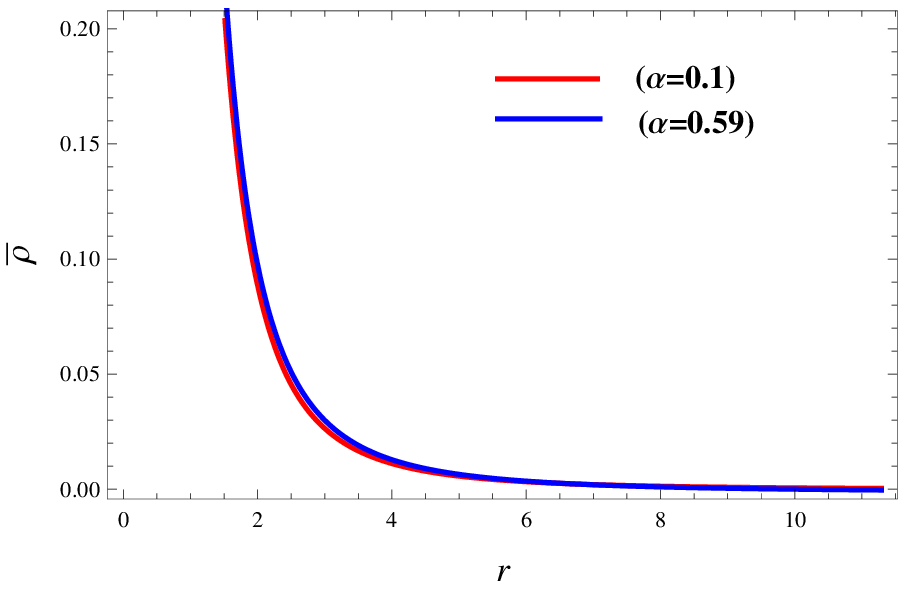,width=0.45\linewidth}\epsfig{file=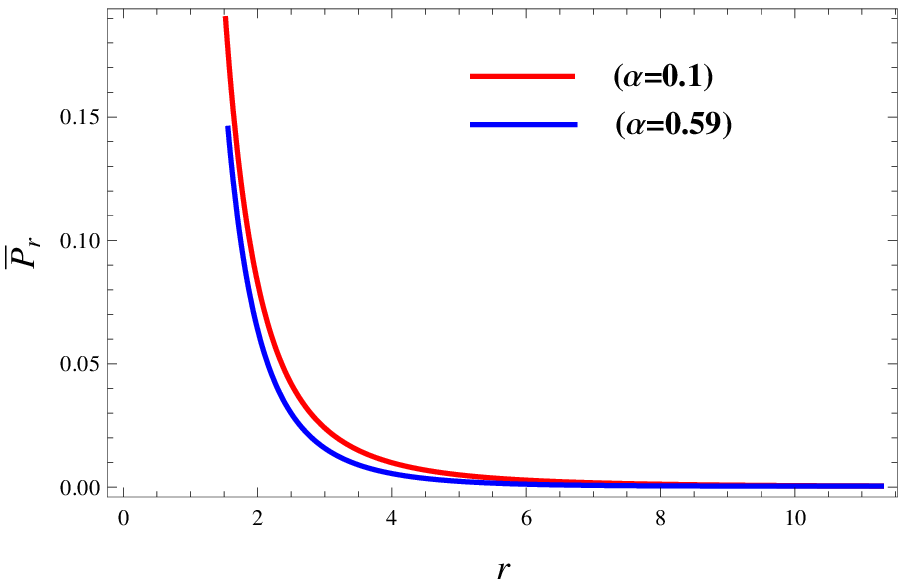,width=0.45\linewidth}
\epsfig{file=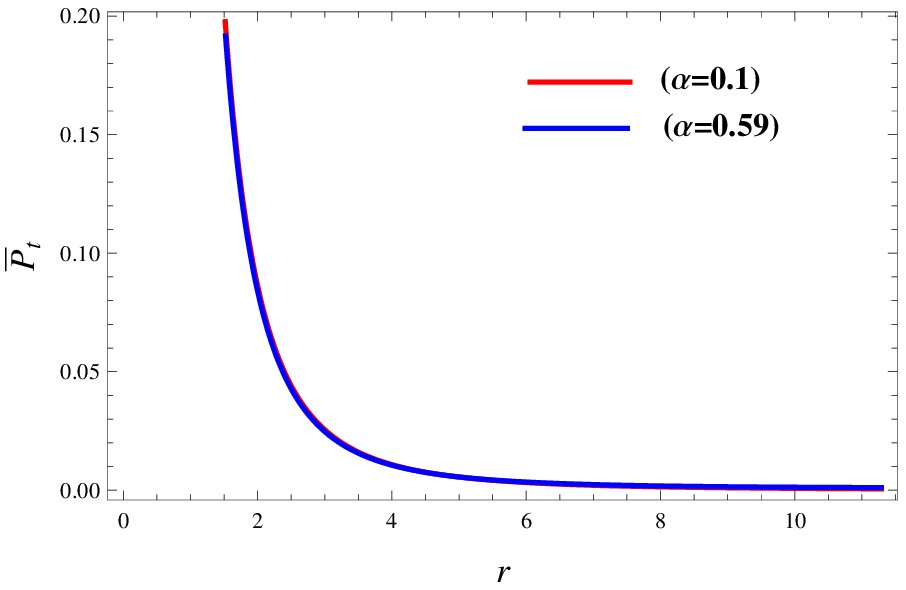,width=0.45\linewidth}\epsfig{file=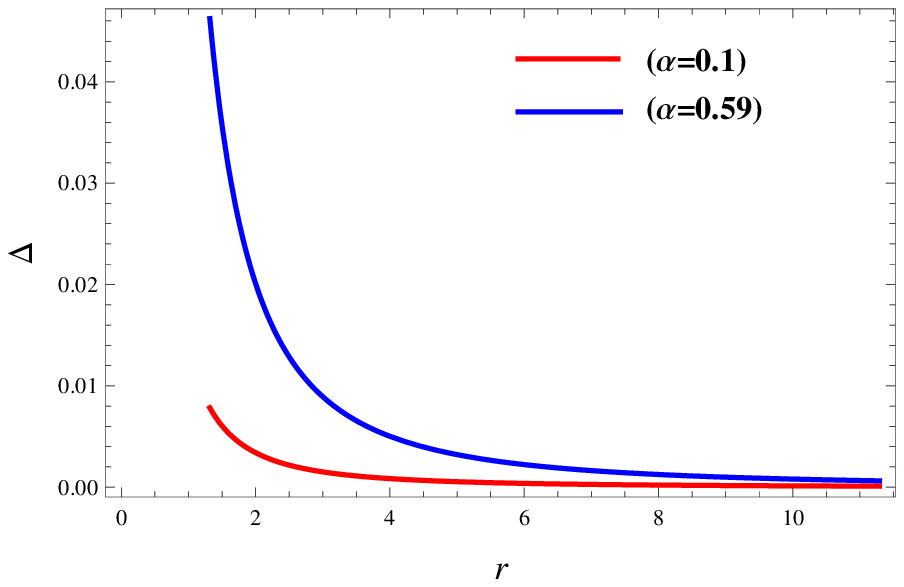,width=0.45\linewidth}
\caption{Behavior of energy density, radial/tangential pressures and
anisotropy of the second solution.}
\end{figure}

\subsection{Solution III}

In order to formulate the third anisotropic solution, we apply a
linear equation of state on $\Phi_{\zeta\nu}$ as \cite{607}
\begin{equation}\label{756}
\Phi^{0}_{0}=\Phi^{1}_{1},
\end{equation}
which leads to
\begin{equation}\label{76}
h(r) = r^{2 n} A,
\end{equation}
where $A$ is an integration constant. By using Eq.(\ref{76}), we
find the anisotropic solution as
\begin{eqnarray}\label{435}
\bar{\rho}&=&\frac{48m\pi^{2}r^{3}-8(1+2n)\pi^{2}r^{4+2n}A\alpha-27m^{2}\beta}{64\pi^{3}r^{6}},\\\label{345}
\bar{P_{r}}&=&\frac{48m\pi^{2}r^{3}+8(1+2n)\pi^{2}r^{4+2n}A\alpha-
9m^{2}\beta}{64\pi^{3}r^{6}},\\\label{99099}
\bar{P_{t}}&=&\frac{48m\pi^{2}r^{3}+8n(1+2n)\pi^{2}r^{4+2n}A\alpha-
9m^{2}\beta}{64\pi^{3}r^{6}}.
\end{eqnarray}
We consider the same values of $M$, $\mathcal{R}$ and $\beta$ as in
solution I and II. Moreover, $A$ is fixed as $-0.02$ while we take
two values of the decoupling parameter as $0.1$ and $0.9$. From
Figure \textbf{3}, it is observed that $\bar\rho,~\bar P_{r},~\bar
P_{t}$ and anisotropy have the maximum values near the core of the
star and decrease continuously as $r$ increases.
\begin{figure}\center
\epsfig{file=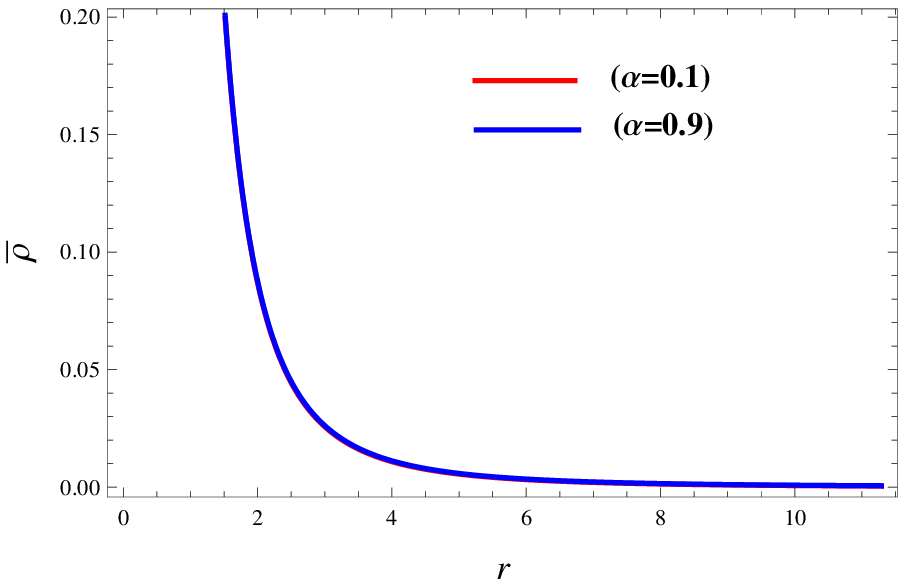,width=0.45\linewidth}\epsfig{file=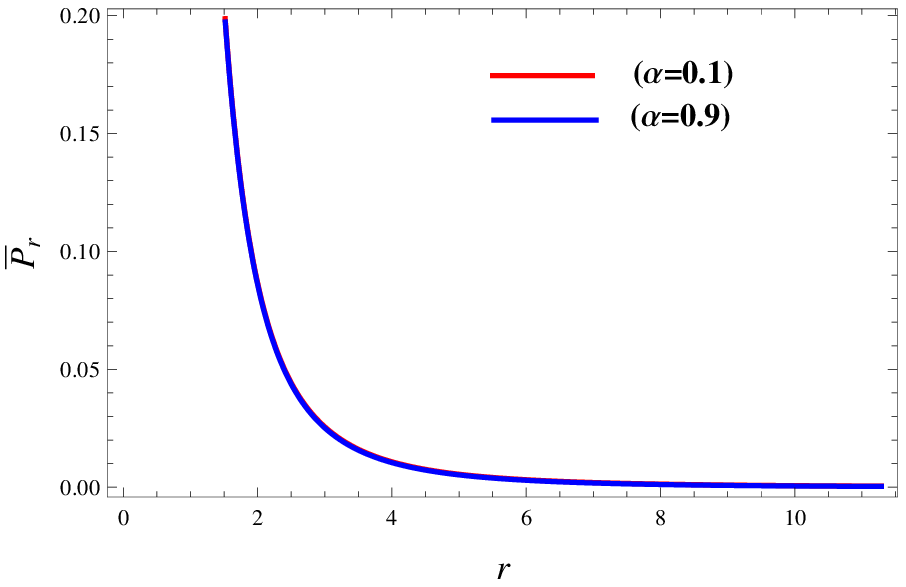,width=0.45\linewidth}
\epsfig{file=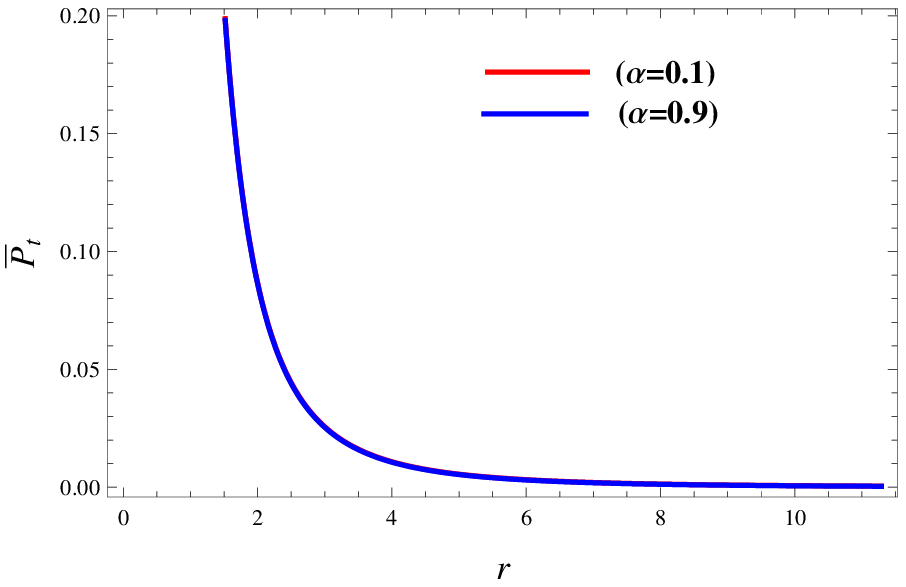,width=0.45\linewidth}\epsfig{file=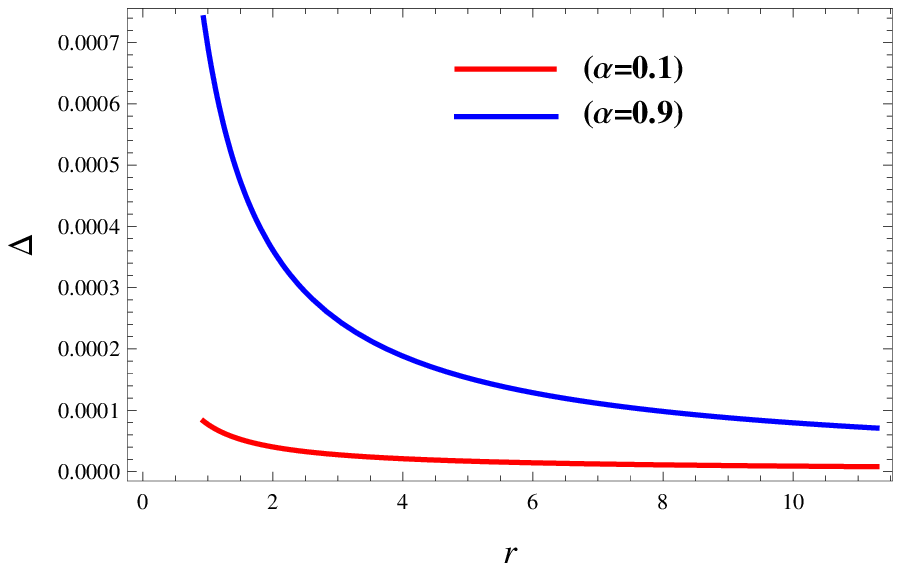,width=0.45\linewidth}
\caption{Behavior of energy density, radial/tangential pressures and
anisotropy of the third solution.}
\end{figure}

\section{Graphical Analysis}

We check the essential physical characteristics of the anisotropic
solutions using energy conditions, stability criteria, mass,
compactness and redshift parameter.

\subsection{Energy Conditions}

Energy constraints are used to check the existence of ordinary
matter as well as physical viability of the resulting solutions.
These constraints are classified into weak, strong, null and
dominant energy conditions. For anisotropic matter distribution,
these constraints are defined as
\begin{eqnarray}\nonumber
&&WEC:~\bar{\rho}+\bar{P_{r}}\geq0, \quad \bar{\rho}\geq0, \quad
\bar{\rho}+\bar{P_{t}}\geq0,
\\\nonumber
&&SEC:~\bar{\rho}+\bar{P_{t}}\geq0,\quad,
\bar{\rho}+\bar{P_{r}}\geq0,\quad
\bar{\rho}+\bar{P_{r}}+2\bar{P_{t}}\geq0, \\\nonumber
&&NEC:~\bar{\rho}+\bar{P_{r}}\geq0,\quad
\bar{\rho}+\bar{P_{t}}\geq0,
\\\nonumber
&&DEC:~\bar{\rho}-|\bar{P_{r}}|\geq0,\quad
\bar{\rho}-|\bar{P_{t}}|\geq0.
\end{eqnarray}
The energy conditions show the same trend as the matter variables
and all solutions are consistent with the criteria of physically
acceptable solutions for the assumed values of parameters.
\begin{figure}\center
\epsfig{file=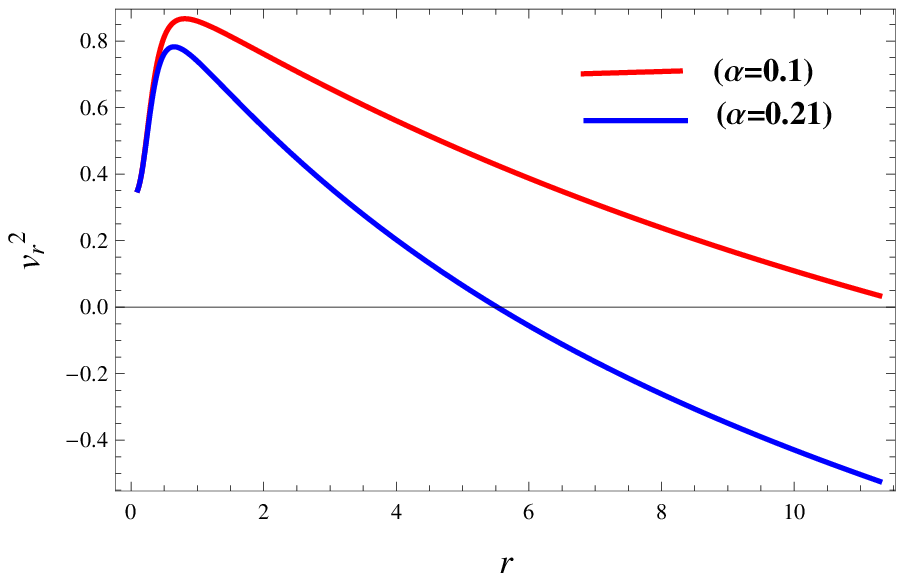,width=0.45\linewidth}\epsfig{file=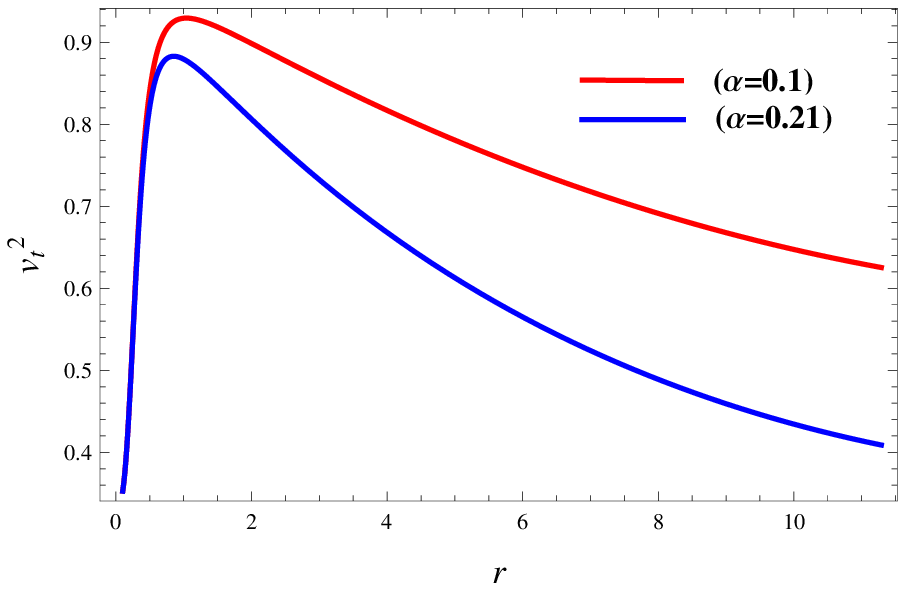,width=0.45\linewidth}
\epsfig{file=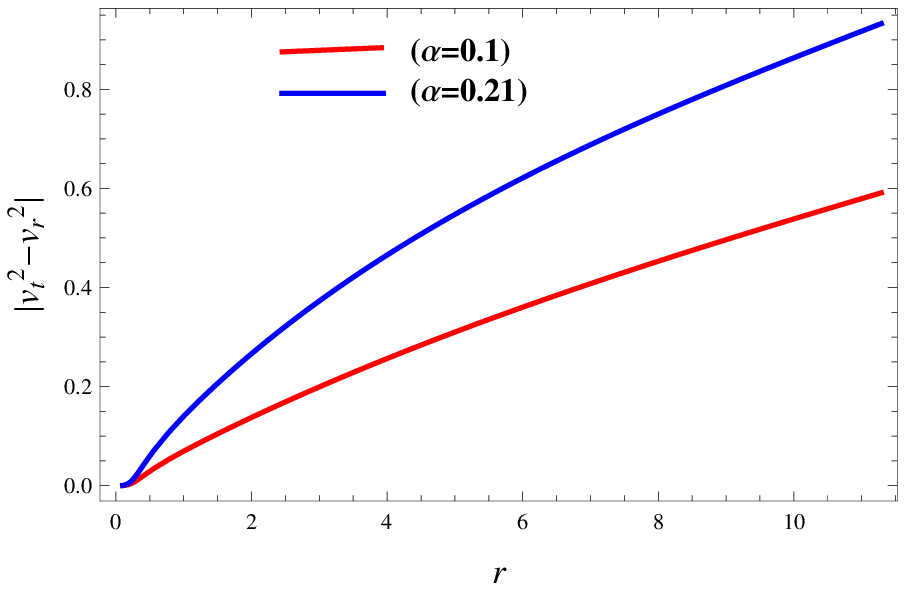,width=0.45\linewidth} \caption{Plots of
causality condition and Herrera's cracking approach of solution I.}
\end{figure}

\subsection{Causality Condition}

A compact structure is stable if it satisfies the condition that the
speed of light exceeds the speed of sound traveling through its
medium \cite{992}. Thus, the radial and tangential components of
sound velocity should, respectively, satisfy the inequalities
\begin{equation}\label{81**}
0 < v^{2}_{r} < 1, \quad 0 < v^{2}_{t} < 1,
\end{equation}
where $v^{2}_{r}=\frac{d\bar{P_{r}}}{d\bar{\rho}}$ and
$v^{2}_{t}=\frac{d\bar{P_{t}}}{d\bar{\rho}}$. The speed of sound in
radial direction violates the causality limit for $\alpha=0.21$ and
$0.59$ corresponding to solution I and II, respectively. The speed
of sound in tangential direction corresponding to first, second and
third anisotropic model satisfies the stability limit as shown in
Figures \textbf{4}-\textbf{6}. Herrera's cracking approach
\cite{997} is another procedure to check the stability of models.
The solutions are stable if the speed of sound in the radial and
tangential directions satisfy the limit $0<|v^{2}_{t}-v^{2}_{r}|<1$.
The anisotropic Tolman V solutions satisfy this limit as shown in
Figures \textbf{4}-\textbf{6}.
\begin{figure}\center
\epsfig{file=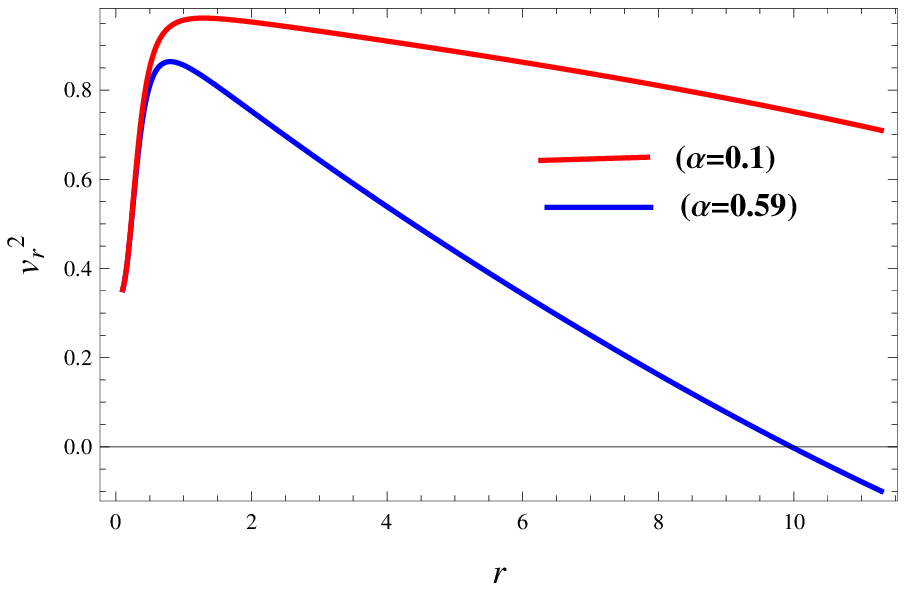,width=0.45\linewidth}\epsfig{file=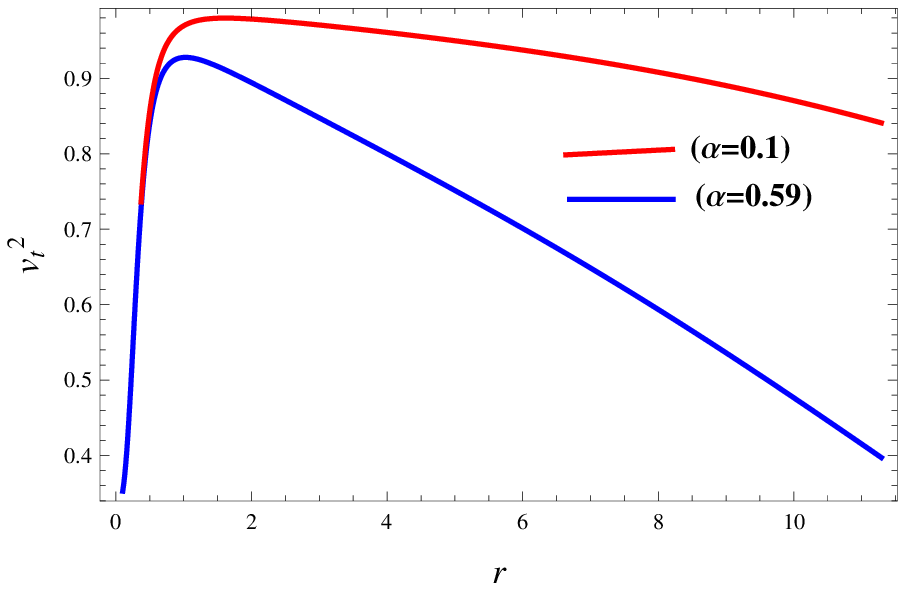,width=0.45\linewidth}
\epsfig{file=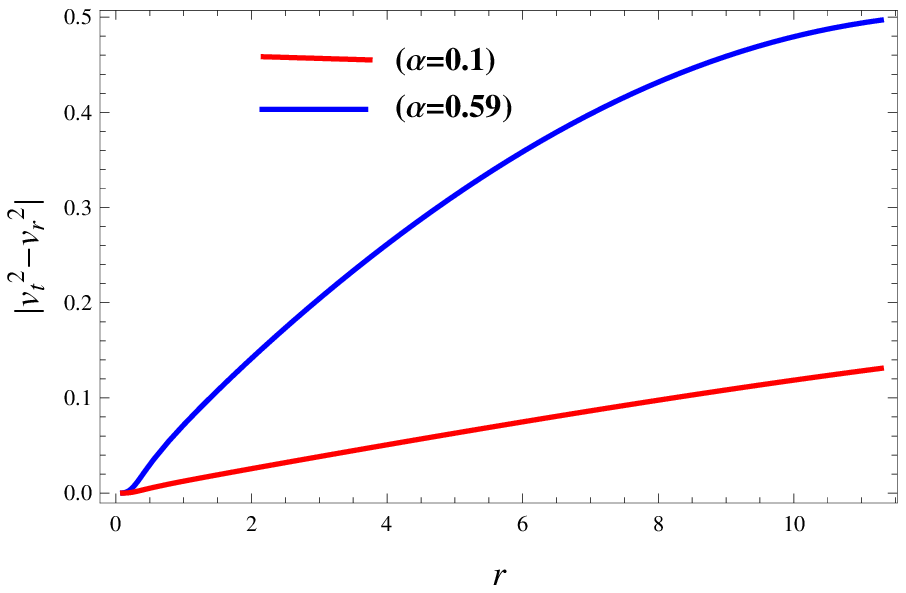,width=0.45\linewidth} \caption{Plots of
causality condition and Herrera's cracking approach of solution II.}
\end{figure}
\begin{figure}\center
\epsfig{file=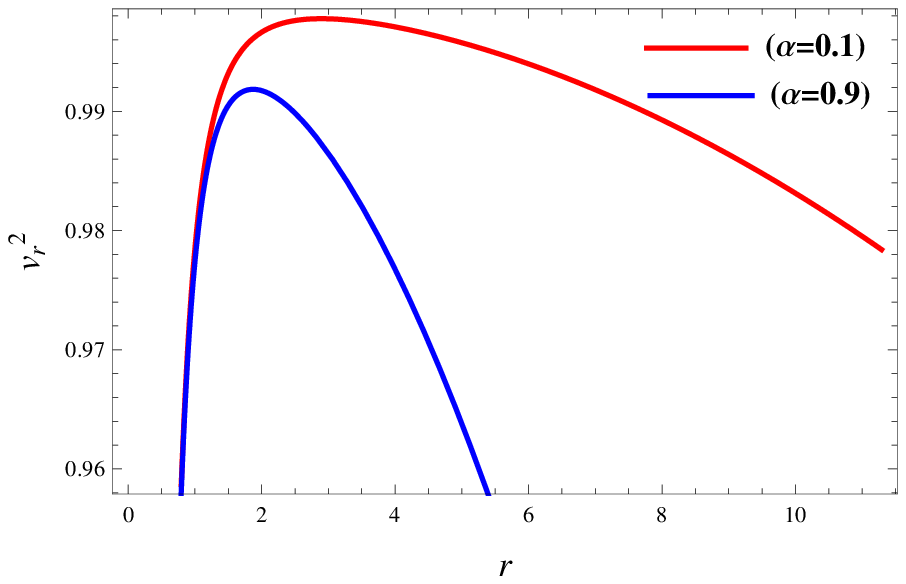,width=0.45\linewidth}\epsfig{file=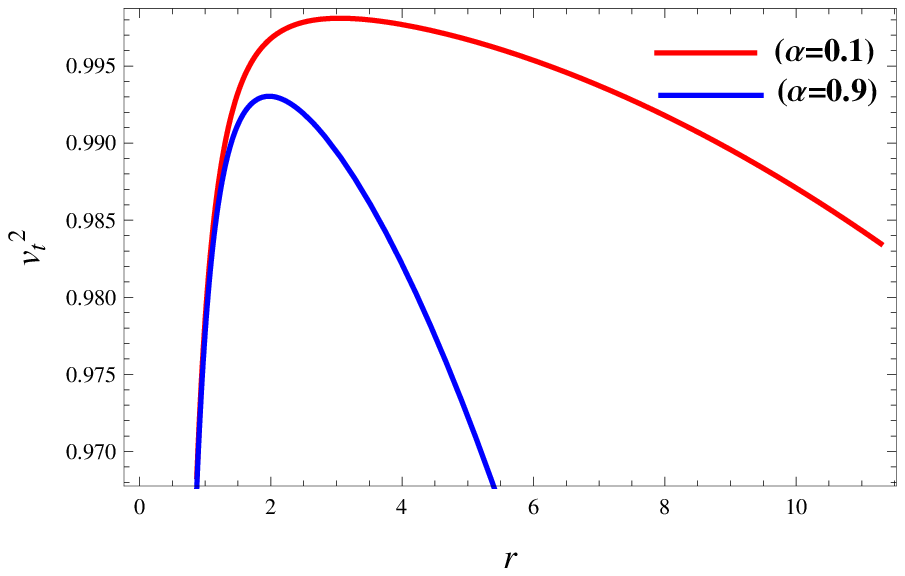,width=0.45\linewidth}
\epsfig{file=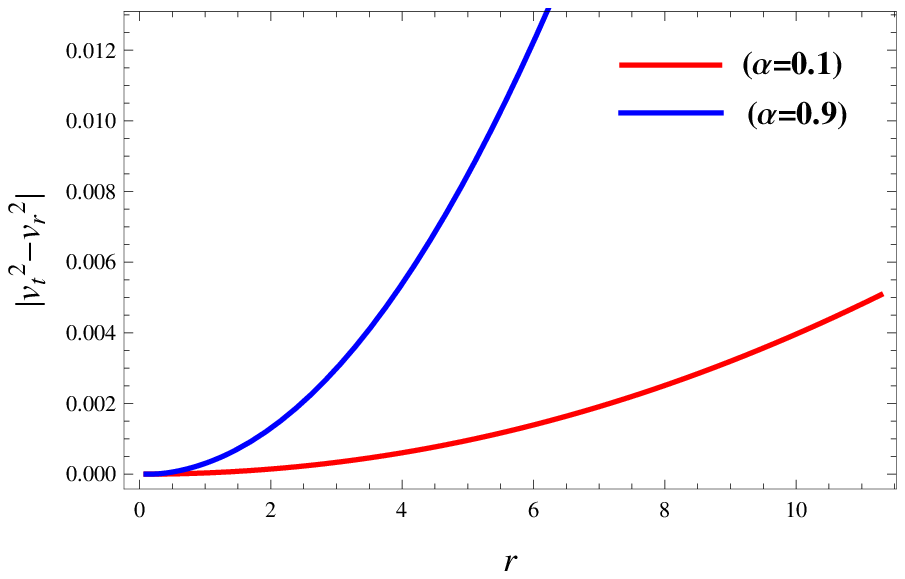,width=0.45\linewidth} \caption{Plots of
causality condition and Herrera's cracking approach of solution
III.}
\end{figure}

\subsection{Redshift and Compactness}

The effective mass of self-gravitating object for anisotropic
spherically symmetric fluid distribution is given as
\begin{equation}\label{85}
m(r)=4\pi\int_{0}^{r}{{\bar{\rho}}r^{2}}dr.
\end{equation}
It is very difficult to find effective mass by using the above
relation in the modified theories. Therefore, we use an effective
way to find the mass using the metric components as
\begin{equation}\label{88}
m(r) = \frac{r(1-e^{-\eta(r)})}{2}.
\end{equation}
The mass of the above three solutions increases as r increases as
shown in Figures \textbf{7}-\textbf{9}. The compactness of the
stellar system is determined by using the relation
\begin{figure}\center
\epsfig{file=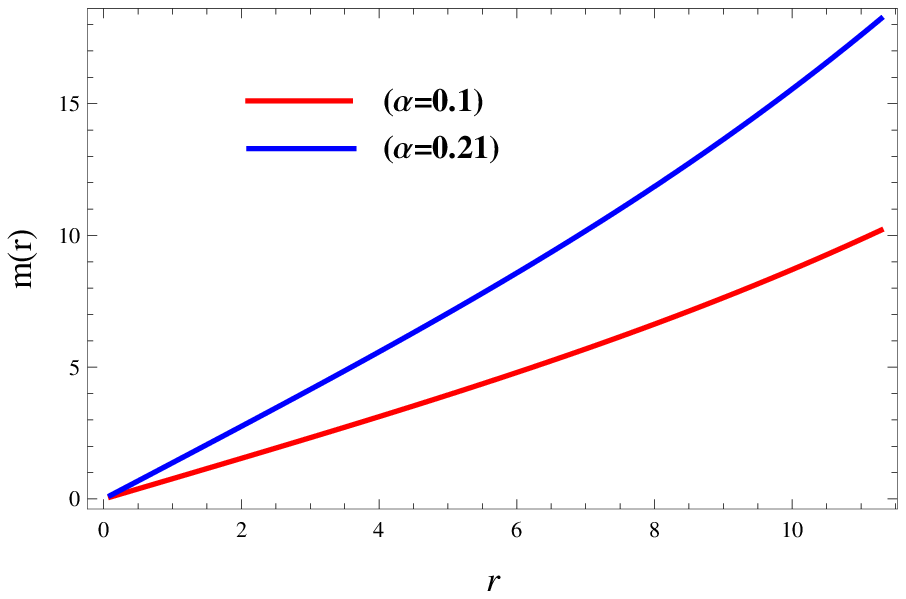,width=0.45\linewidth}\epsfig{file=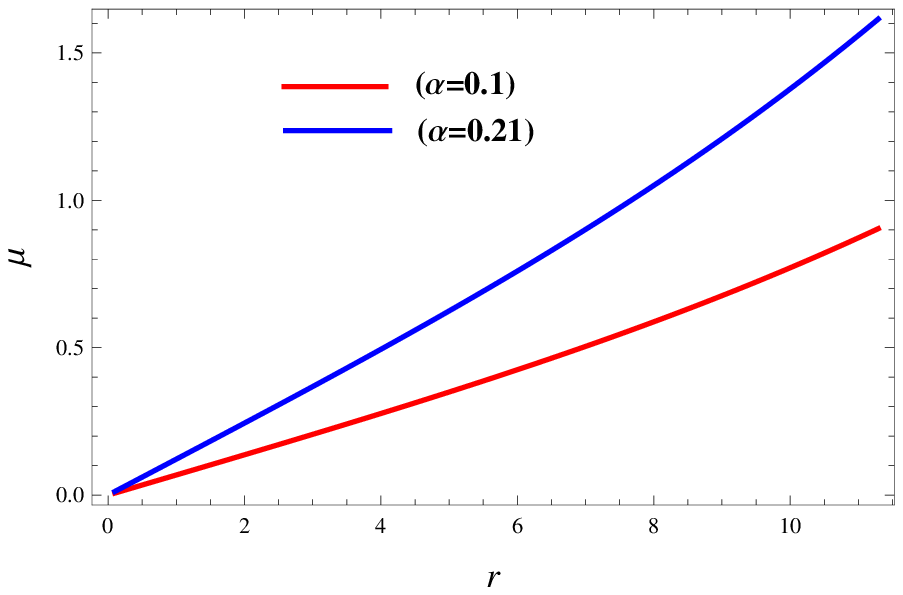,width=0.45\linewidth}
\epsfig{file=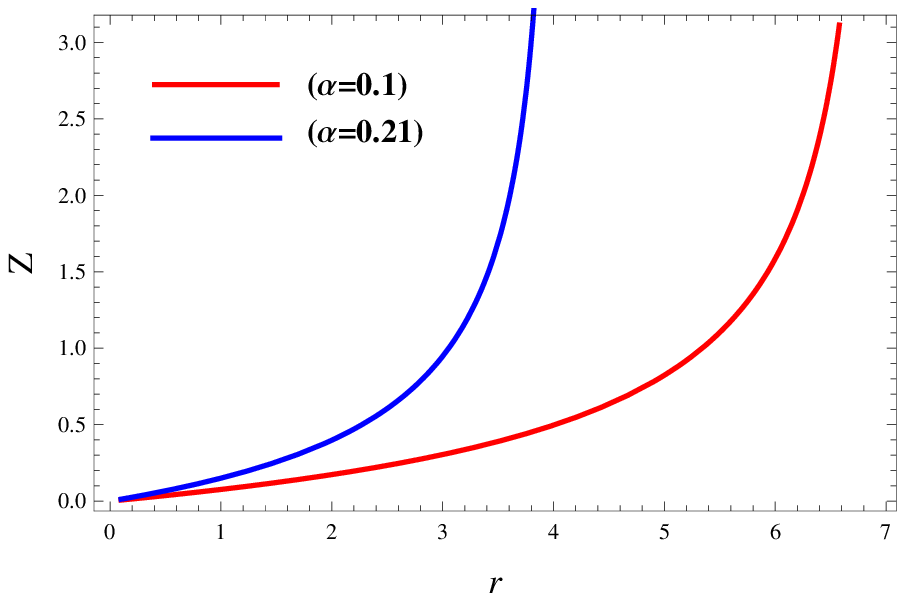,width=0.45\linewidth} \caption{Behavior of mass,
compactness and redshift of solution I.}
\end{figure}
\begin{figure}\center
\epsfig{file=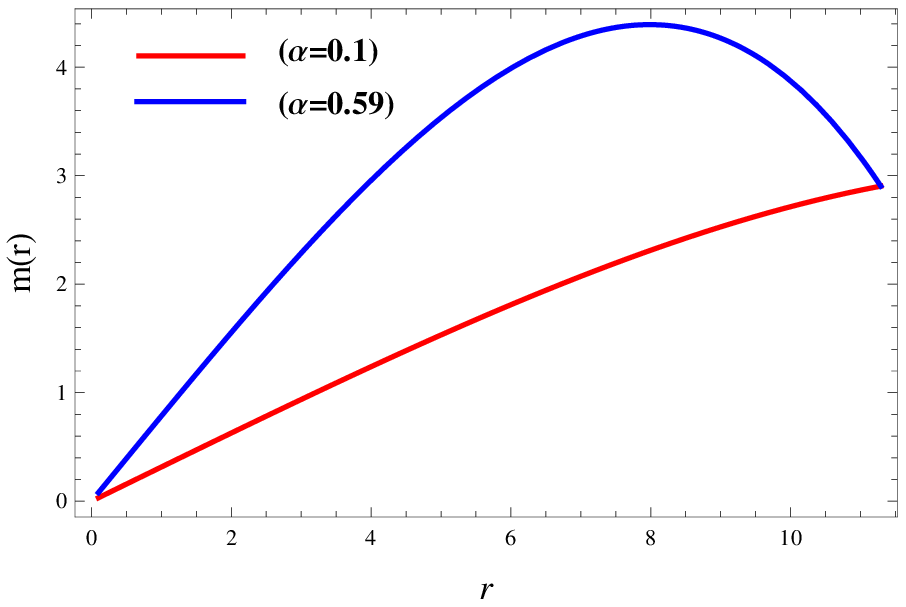,width=0.45\linewidth}\epsfig{file=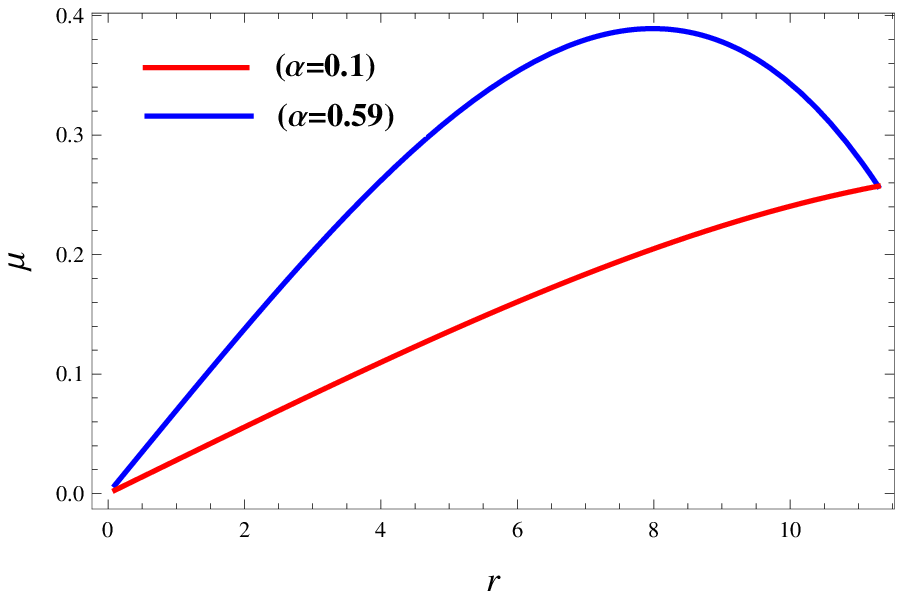,width=0.45\linewidth}
\epsfig{file=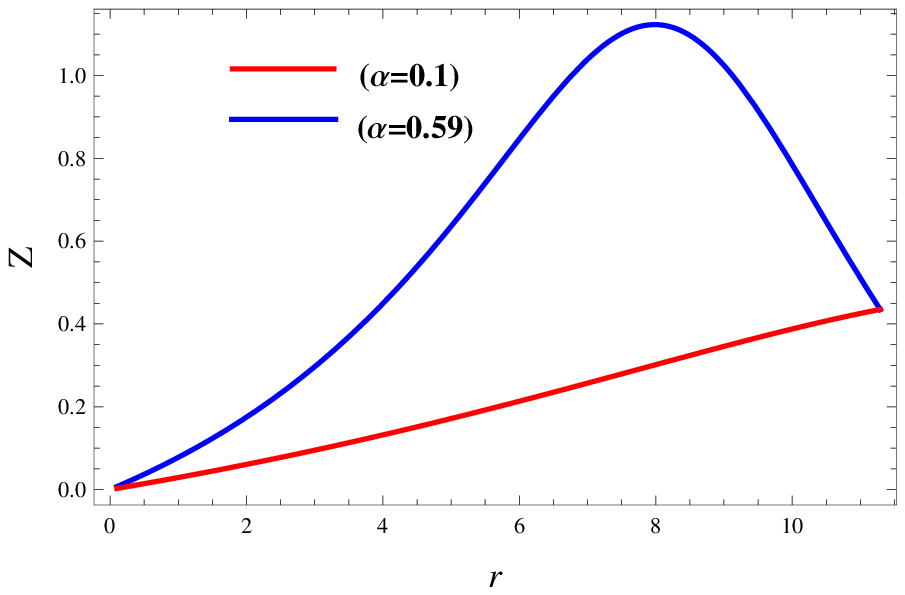,width=0.45\linewidth} \caption{Plots of mass,
compactness and redshift of solution II.}
\end{figure}
\begin{figure}\center
\epsfig{file=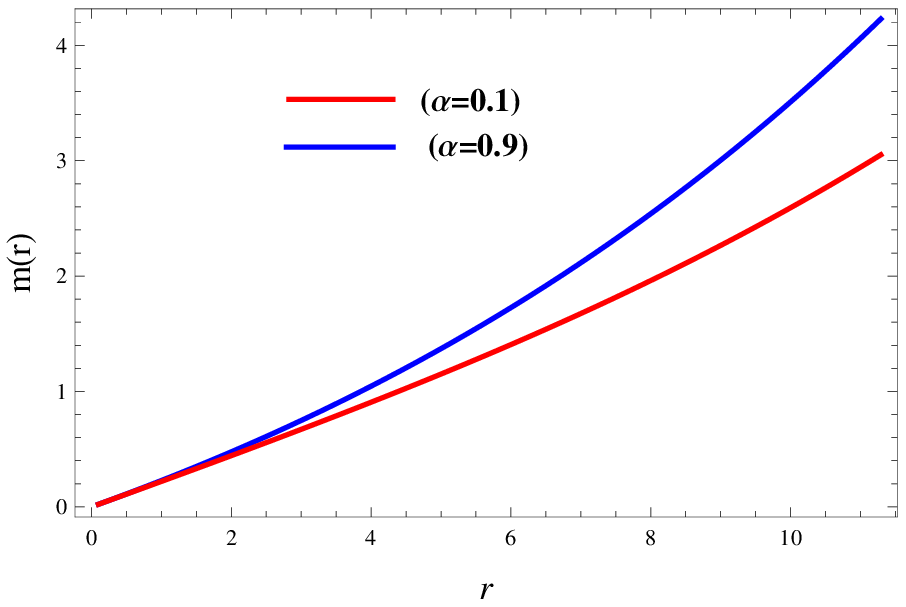,width=0.45\linewidth}\epsfig{file=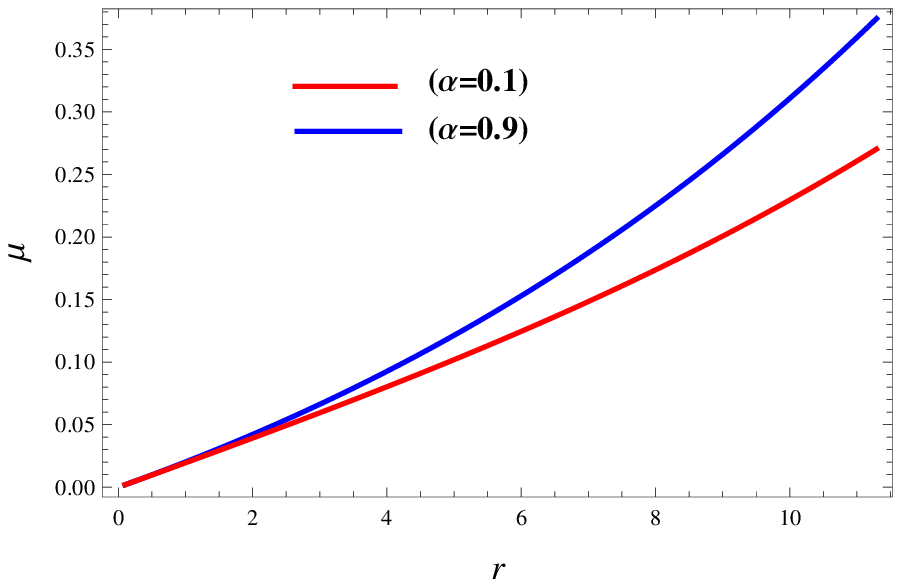,width=0.45\linewidth}
\epsfig{file=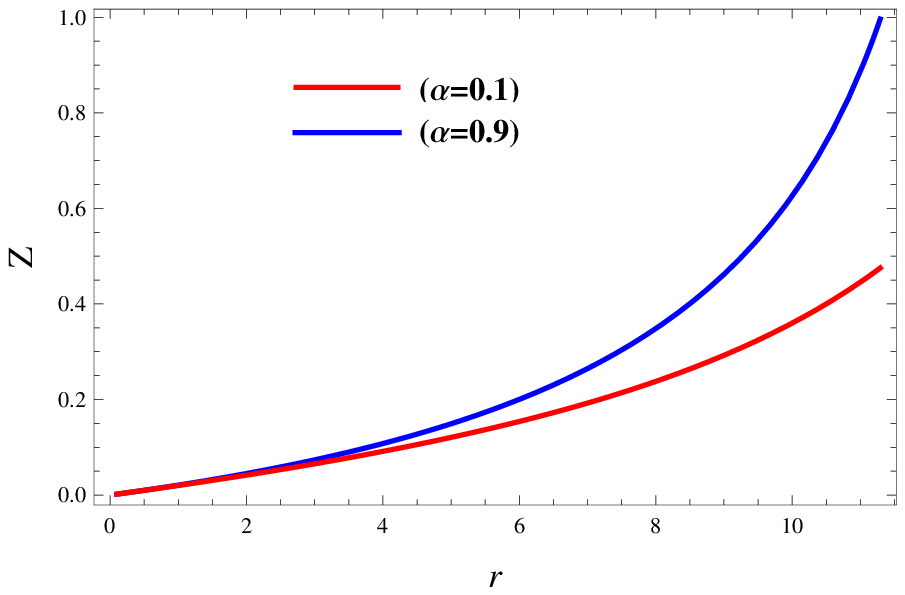,width=0.45\linewidth} \caption{Behavior of mass,
compactness and redshift of solution III.}
\end{figure}
\begin{equation}\label{88}
\mu(r)=\frac{m(r)}{r}.
\end{equation}
The upper limit of compactness for anisotropic fluid
$(\frac{M}{\mathcal{R}}<\frac{4}{9})$ was measured by Buchdahl
\cite{990}. The compactness parameter of solution I does not attain
the corresponding limit (Figure \textbf{7}). Second and third
anisotropic Tolman V solutions agree with this limit (Figures
\textbf{8} and \textbf{9}). The surface redshift of self-gravitating
body measures the increment in wavelength of electromagnetic
radiation due to the gravitational pull of that body. The redshift
parameter is defined as
\begin{equation}\label{89}
Z=-1+\frac{1}{\sqrt{1-2\mu}},
\end{equation}
whose maximum value for anisotropic solution is $Z\leq4.77$
\cite{991}. The graph of redshift parameter increases as $r$
increases for all the anisotropic solutions, as shown in Figures
\textbf{7}-\textbf{9}. For solutions II and III, the values of
redshift parameter remains below the value $4.77$ for the chosen
values of the model and decoupling parameters. The redshift
parameter corresponding to the solution I does not attain the
corresponding limit.

\section{Concluding Remarks}

Gravitational decoupling through the MGD method has been extensively
used to obtain well-behaved models of the self-gravitating system.
In this paper, we have constructed anisotropic solutions using this
approach in the context of EMSG. For this reason, we have added a
new anisotropic source alongside the isotropic energy-momentum
tensor. The field equations containing anisotropic as well as seed
matter distributions have been decoupled into two sets through a
radial transformation. One set consists of the standard field
equations of EMSG while the other corresponds to the anisotropic
source. To check the effectiveness of this technique, we have
described the first array by considering the metric potentials of
the isotropic Tolman V solution along with the model
$f(R,T^{2})=R+\beta T^{2}$. We have applied the junction conditions
by taking the Schwarzschild metric as exterior spacetime to find the
values of constants appearing in the solution. The physical
viability of all solutions is analyzed through the behavior of
effective parameters, anisotropy and energy constraints for some
specific values of decoupling and model parameters. We have also
checked the stability of the extended solutions through causality
condition and Herrera's cracking approach. The behavior of mass,
compactness and redshift of anisotropic extensions has also been
investigated.

For the first solution, we have applied the constraint $\Phi^{0}_{0}
=\tilde\rho$, whereas the second and third solutions have been
constructed by using a constraint $\Phi^{1}_{1}=\tilde P$ and a
linear equation of state relating $\Phi^{0}_{0}$ and $\Phi^{1}_{1}$,
respectively. We have graphically examined the new anisotropic
solutions by utilizing the radius and mass of PSR J1614-2230 star.
Further, the values of parameters are fixed as $\beta=-0.1$ with
$\alpha=0.1,~0.21$, $\alpha=0.1,~59$ and $\alpha=0.1,~0.9$ for
solutions I, II and III, respectively. For all solutions, the
effective energy density, radial and tangential pressures attain the
maximum values near the core of star and decrease continuously
towards its boundary. Furthermore, all the obtained solutions are
viable as they satisfy all the energy conditions for the chosen
values of the parameters. The speed of sound in radial direction
does not satisfy the stability limit for $\alpha=0.21$ and $0.59$
for solution I and II, respectively whereas the solution III
satisfies this criteria for $\alpha=0.1,~0.9$. Moreover, the
Herrera's cracking approach lies within the limit
$0<|v^{2}_{t}-v^{2}_{r}|<1$ for all anisotropic solutions. The mass,
compactness and redshift show increasing behavior for above three
solutions and the limits $\frac{M}{\mathcal{R}}<\frac{4}{9}$ and
$Z\leq4.77$ are fulfilled only by solutions II and III. We conclude
that solution III is physically viable and stable in the framework
of EMSG for all values of the decoupling parameter, whereas
solutions I and II are viable only for $\alpha\leq0.21$ and
$\alpha\leq0.59$, respectively.

\section*{Appendix A}
\renewcommand{\theequation}{A\arabic{equation}}
\setcounter{equation}{0}

The energy conditions corresponding to solution I are
\begin{eqnarray}\nonumber
\bar{\rho}&=&\frac{\Bigg(n^{2}
-2n+\left(\frac{r}{F}\right)^{W}(1+W)(n^{2}-1-2n)\Bigg)\alpha}{(n^{2}-1-2n)r^{2}}\\\nonumber&+&\frac{3m(16\pi^{2}r^{3}
-9m\beta)}{64\pi^{3}r^{6}}, \\\nonumber
\bar{\rho}+\bar{P_{r}}&=&\frac{\Bigg(\left(\frac{r}{F}\right)^{W}(1+W)(n^{2}-1-2n)
-2n+n^{2}\Bigg)\alpha}{(n^{2}-1-2n)r^{2}}\\\nonumber&-&\frac{(1 +
2n)\Bigg(n^{2}-2n+\left(\frac{r}{F}\right)^{W}(n^{2}-1-2n)\Bigg)\alpha}{(n^{2}
-1-2n)r^{2}}\\\nonumber&+&\frac{48m\pi^{2}r^{3}-9m^{2}\beta}{64\pi^{3}r^{6}}+\frac{3m(16\pi^{2}r^{3}
-9m\beta)}{64\pi^{3}r^{6}}, \\\nonumber
\bar{\rho}+\bar{P_{t}}&=&\frac{32r^{4}\left(\frac{r}{F}\right)^{W}W(1+3n+n^{2}-
n^{3})-\left(\frac{r}{F}\right)^{W}(64r^{4}n^{4}-128r^{4}n^{3})}
{64r^{6}(n^{2}-2n-1)}\alpha\\\nonumber&+&\frac{64r^{4}n^{2\left(\frac{r}{F}\right)^{W}
}-64r^{4}n^{4}+128r^{4}n^{3}}{64r^{6}
(n^{2}-2n-1)}\alpha+\frac{48m\pi^{2}
r^{3}-9m^{2}\beta}{64r^{6}\pi^{3}}\\\nonumber&+&\frac{\Bigg(\left(\frac{r}{F}\right)^{W}(W+1)(n^{2}-1-2n)-2n+n^{2}
\Bigg)\alpha}{(n^{2}-1-2n)r^{2}}\\\nonumber&+&\frac{3m(16\pi^{2}r^{3}
-9m\beta)}{64\pi^{3}r^{6}}, \\\nonumber
\bar{\rho}+2\bar{P_{t}}+\bar{P_{r}}&=&2\Bigg(\frac{32r^{4}\left(\frac{r}{F}\right)^{W}W(1+3n+n^{2}-
n^{3})-\left(\frac{r}{F}\right)^{W}(64r^{4}n^{4}-128r^{4}n^{3})}
{64r^{6}(n^{2}-2n-1)}\alpha\\\nonumber&+&\frac{64r^{4}n^{2\left(\frac{r}{F}\right)^{W}
}-64r^{4}n^{4}+128r^{4}n^{3}}{64r^{6}
(n^{2}-2n-1)}\alpha+\frac{48m\pi^{2}
r^{3}-9m^{2}\beta}{64r^{6}\pi^{3}}\Bigg)\\\nonumber&+&\frac{\Bigg(\left(\frac{r}{F}\right)^{W}(1+W)(n^{2}-2n-1)-2n+n^{2}
\Bigg)\alpha}{(n^{2}-1-2n)r^{2}}\\\nonumber&-&\frac{(1 +
2n)\Bigg(\left(\frac{r}{F}\right)^{W}(n^{2}-1-2n)+n^{2}-2n\Bigg)\alpha}{(n^{2}
-2n-1)r^{2}}\\\nonumber &+&\frac{3m(16\pi^{2}r^{3}
-9m\beta)}{64\pi^{3}r^{6}}+\frac{48m\pi^{2}r^{3}-9m^{2}\beta}{64\pi^{3}r^{6}},
\\\nonumber
\bar{\rho}-\bar{P_{r}}&=&\frac{\Bigg(\left(\frac{r}{F}\right)^{W}(1+W)(n^{2}-2n-1)-2n+n^{2}
\Bigg)\alpha}{(n^{2}-1-2n)r^{2}}\\\nonumber&+&-\frac{(1 +
2n)\Bigg(\left(\frac{r}{F}\right)^{W}(n^{2}-1-2n)+n^{2}-2n\Bigg)\alpha}{(n^{2}
-2n-1)r^{2}}\\\nonumber&+&\frac{3m(16\pi^{2}r^{3}
-9m\beta)}{64\pi^{3}r^{6}}-\frac{48m\pi^{2}r^{3}-9m^{2}\beta}{64\pi^{3}r^{6}},
\\\nonumber
\bar{\rho}-\bar{P_{t}}&=&-\frac{32r^{4}\left(\frac{r}{F}\right)^{W}W(1+3n+n^{2}-
n^{3})-\left(\frac{r}{F}\right)^{W}(64r^{4}n^{4}-128r^{4}n^{3})}
{64r^{6}(n^{2}-2n-1)}\alpha\\\nonumber&-&\frac{64r^{4}n^{2\left(\frac{r}{F}\right)^{W}
}-64r^{4}n^{4}+128r^{4}n^{3}}{64r^{6}
(n^{2}-2n-1)}\alpha-\frac{48m\pi^{2}
r^{3}-9m^{2}\beta}{64r^{6}\pi^{3}}\\\nonumber&+&\frac{\Bigg(\left(\frac{r}{F}\right)^{W}(1+W)(n^{2}-1-2n)-2n+n^{2}
\Bigg)\alpha}{(n^{2}-1-2n)r^{2}}\\\nonumber&+&\frac{3m(16\pi^{2}r^{3}
-9m\beta)}{64\pi^{3}r^{6}},
\end{eqnarray}
and also corresponding to solution II are
\begin{eqnarray}\nonumber
\bar{\rho}&=&\frac{\left(\frac{r}{F}\right)^{W}(1+W)(1+4n+3n^{2}-2n^{3})-n^{2}}
{(1+2n)(n^{2}-2n-1)r^{2}}\alpha\\\nonumber&+&\frac{3m(16\pi^{2}r^{3}-9m\beta)}{64\pi^{3}r^{6}},
\\\nonumber
\bar{\rho}+\bar{P_{r}}&=&\frac{\left(\frac{r}{F}\right)^{W}(1+4n+3n^{2}-2n^{3})-n^{2}}
{(n^{2}-2n-1)r^{2}}\alpha+\frac{48m\pi^{2}r^{3}-9m^{2}\beta}{64\pi^{3}r^{6}}\\\nonumber&+&
\frac{\left(\frac{r}{F}\right)^{W}(1+W)(1+4n+3n^{2}-2n^{3})-n^{2}}
{(1+2n)(n^{2}-2n-1)r^{2}}\alpha\\\nonumber&+&\frac{3m(16\pi^{2}r^{3}-9m\beta)}
{64\pi^{3}r^{6}}, \\\nonumber
\bar{\rho}+\bar{P_{t}}&=&-\frac{32r^{4}(2n^{2}+8n^{3}+6n^{4}-4n^{5})\left(\frac{r}{F}\right)^{W}
+(1+5n+7n^{2})\left(\frac{r}{F}\right)^{W}W}{64r^{6}(2n+1)(n^{2}-2n-1)}\alpha
\\\nonumber&+&\frac{(n^{3}-2n^{4})\left(\frac{r}{F}\right)^{W}W}{64r^{6}(1+2n)(n^{2}-1-2n)}\alpha
-\frac{2n^{4}}{64r^{6}(1+2n)(n^{2}-2n-1)}\alpha\\\nonumber&+&\frac{\left(\frac{r}
{F}\right)^{W}(1+W)(1+4n+3n^{2}-2n^{3})-n^{2}}
{(1+2n)(n^{2}-2n-1)r^{2}}\alpha\\\nonumber&+&\frac{3m(16\pi^{2}r^{3}-9m\beta)
}{64\pi^{3}r^{6}}+\frac{48m\pi^{2}r^{3}
-9m^{2}\beta}{64r^{6}\pi^{3}}, \\\nonumber
\bar{\rho}+\bar{P_{r}}+2\bar{P_{t}}&=&2
\Bigg(-\frac{32r^{4}(2n^{2}+8n^{3}+6n^{4}-4n^{5})\left(\frac{r}{F}\right)^{W}
+(1+5n+7n^{2})\left(\frac{r}{F}\right)^{W}W}{64r^{6}(1+2n)(n^{2}-2n-1)}\alpha
\\\nonumber&+&\frac{(n^{3}-2n^{4})\left(\frac{r}{F}\right)^{W}W}{64r^{6}(1+2n)(n^{2}-1-2n)}\alpha
-\frac{2n^{4}}{64r^{6}(1+2n)(n^{2}-2n-1)}\alpha\\\nonumber&+&\frac{48m\pi^{2}r^{3}
-9m^{2}\beta}{64r^{6}\pi^{3}}\Bigg)+\frac{\left(\frac{r}{F}\right)^{W}(1+W)(1+4n+3n^{2}-2n^{3})-n^{2}}
{(1+2n)(n^{2}-2n-1)r^{2}}\alpha
\\\nonumber&+&\frac{\left(\frac{r}{F}\right)^{W}(1+4n+3n^{2}-2n^{3})-n^{2}}
{(n^{2}-2n-1)r^{2}}\alpha+\frac{48m\pi^{2}r^{3}-9m^{2}\beta}{64\pi^{3}r^{6}}\\\nonumber&+&
\frac{3m(16\pi^{2}r^{3}-9m\beta)}{64\pi^{3}r^{6}},
\\\nonumber
\bar{\rho}-\bar{P_{r}}&=&\frac{\left(\frac{r}{F}\right)^{W}(1+W)(1+4n+3n^{2}-2n^{3})-n^{2}}
{(1+2n)(n^{2}-2n-1)r^{2}}\alpha+\frac{3m(16\pi^{2}r^{3}+15m\beta)}{64\pi^{3}r^{6}}
\\\nonumber&-&\frac{\left(\frac{r}{F}\right)^{W}(1+4n+3n^{2}-2n^{3})-n^{2}}
{(n^{2}-2n-1)r^{2}}\alpha-\frac{48m\pi^{2}r^{3}-9m^{2}\beta}{64\pi^{3}r^{6}},
\\\nonumber
\bar{\rho}-\bar{P_{t}}&=&\frac{32r^{4}(2n^{2}+8n^{3}+6n^{4}-4n^{5})\left(\frac{r}{F}\right)^{W}
+(1+5n+7n^{2})\left(\frac{r}{F}\right)^{W}W}{64r^{6}(2n+1)(n^{2}-1-2n)}\alpha
\\\nonumber&-&\frac{(n^{3}-2n^{4})W\left(\frac{r}{F}\right)^{W}}{64r^{6}(1+2n)(n^{2}-1-2n)}\alpha
+\frac{2n^{4}}{64r^{6}(1+2n)(n^{2}-2n-1)}\alpha\\\nonumber&-&\frac{48m\pi^{2}r^{3}
-9m^{2}\beta}{64r^{6}\pi^{3}}+\frac{3m(16\pi^{2}r^{3}-9m\beta)}{64\pi^{3}r^{6}}
\\\nonumber&+&\frac{\left(\frac{r}{F}\right)^{W}(1+W)(1+4n+3n^{2}-2n^{3})-n^{2}}
{(1+2n)(n^{2}-2n-1)r^{2}}\alpha.
\end{eqnarray}
These conditions become corresponding to solution III as
\begin{eqnarray}\nonumber
\bar{\rho}&=&\frac{48m\pi^{2}r^{3}-8(1+2n)\pi^{2}r^{4+2n}A\alpha-27m^{2}\beta}{64\pi^{3}r^{6}},\\\nonumber
\bar{\rho}+\bar{P_{r}}&=&\frac{48m\pi^{2}r^{3}-8(1+2n)\pi^{2}r^{4+2n}A\alpha-27m^{2}\beta}
{64\pi^{3}r^{6}}\\\nonumber&+&\frac{48m\pi^{2}r^{3}+8(1+2n)\pi^{2}r^{4+2n}A\alpha-
9m^{2}\beta}{64\pi^{3}r^{6}},\\\nonumber
\bar{\rho}+\bar{P_{t}}&=&\frac{48m\pi^{2}r^{3}-8(1+2n)\pi^{2}r^{4+2n}A\alpha-27m^{2}\beta}
{64\pi^{3}r^{6}}\\\nonumber&+&\frac{48m\pi^{2}r^{3}+8n(1+2n)\pi^{2}r^{4+2n}A\alpha-
9m^{2}\beta}{64\pi^{3}r^{6}},\\\nonumber
\bar{\rho}+\bar{P_{r}}+2\bar{P_{t}}&=&\frac{48m\pi^{2}r^{3}-8(1+2n)\pi^{2}r^{4+2n}A\alpha-27m^{2}\beta}
{64\pi^{3}r^{6}}\\\nonumber&+&\frac{48m\pi^{2}r^{3}+8(1+2n)\pi^{2}r^{4+2n}A\alpha-
9m^{2}\beta}{64\pi^{3}r^{6}}\\\nonumber&+&2\Bigg(\frac{48m\pi^{2}r^{3}+8n(1+2n)\pi^{2}r^{4+2n}A\alpha-
9m^{2}\beta}{64\pi^{3}r^{6}}\Bigg),
\\\nonumber
\bar{\rho}-\bar{P_{r}}&=&\frac{48m\pi^{2}r^{3}-8(1+2n)\pi^{2}r^{4+2n}A\alpha-27m^{2}\beta}{64\pi^{3}
r^{6}}\\\nonumber&-&\frac{48m\pi^{2}r^{3}+8(1+2n)\pi^{2}r^{4+2n}A\alpha-
9m^{2}\beta}{64\pi^{3}r^{6}},
\\\nonumber
\bar{\rho}-\bar{P_{t}}&=&\frac{48m\pi^{2}r^{3}-8(1+2n)\pi^{2}r^{4+2n}A\alpha-27m^{2}\beta}{64\pi^{3}r^{6}}
\\\nonumber&-&\frac{48m\pi^{2}r^{3}+8n(1+2n)\pi^{2}r^{4+2n}A\alpha-
9m^{2}\beta}{64\pi^{3}r^{6}}.
\end{eqnarray}

\end{document}